\documentclass[twocolumn]{aastex631}

\usepackage{natbib}
\usepackage{multirow}
\usepackage{booktabs}
\usepackage{epsfig}
\usepackage{ulem}
\usepackage{rotating}
\usepackage{graphicx}
\usepackage{url}
\usepackage{amsmath}
\usepackage{color}
\usepackage{savesym}
\savesymbol{tablenum}
\usepackage{siunitx}
\usepackage[normalem]{ulem}
\usepackage{appendix}
\usepackage{enumerate}
\usepackage{gensymb}
\usepackage{longtable}
\usepackage{afterpage}
\usepackage{threeparttable}
\usepackage{tabularx}
\usepackage[bottom]{footmisc}

\DeclareSIUnit\parsec{pc}
\DeclareSIUnit\h{\textit{h}}

\DeclareRobustCommand{\uvec}[1]{{%
  \ifcsname uvec#1\endcsname
     \csname uvec#1\endcsname
   \else
    \bm{\hat{\mathbf{#1}}}%
   \fi
}}

\newcommand{\Ydepth}{20.7}
\newcommand{\Ydepthstd}{0.6}
\newcommand{\Jdepth}{20.1}
\newcommand{\Jdepthstd}{0.8}
\newcommand{\Hdepth}{19.3}
\newcommand{\Hdepthstd}{0.6}

\newcommand{\totepochs}{18}
\newcommand{\Yepochs}{6}

\defcitealias{Galbany20kku}{Galbany et al.}
\defcitealias{Galbany20kyx}{Galbany et al.}
\defcitealias{Balcon20mnv}{C.~Balcon}
\defcitealias{Soraisam20nef}{Soraisam et al.}
\defcitealias{Pellegrino20tkp}{Pellegrino et al.}
\defcitealias{Balcon20unl}{C.~Balcon}
\defcitealias{Balcon21zfs}{C.~Balcon}
\defcitealias{Itagaki21fxy}{K.~Itagaki}

\begin{document}
\title{The DEHVILS Survey Overview and Initial Data Release: High-Quality Near-Infrared Type Ia Supernova Light Curves at Low Redshift}
    \author[0000-0001-8596-4746]{Erik R.~Peterson}
    \affiliation{Department of Physics, Duke University, Durham, NC 27708, USA}
    \author[0000-0002-6230-0151]{David O. Jones}
    \affiliation{Gemini Observatory, NSF's NOIRLab, 670 N. A'ohoku Place, Hilo, HI 96720, USA}
    \author[0000-0002-4934-5849]{Daniel Scolnic}
    \affiliation{Department of Physics, Duke University, Durham, NC 27708, USA}
    \author[0000-0002-8687-0669]{Bruno O. Sánchez}
    \affiliation{Department of Physics, Duke University, Durham, NC 27708, USA}
    \author[0000-0003-3429-7845]{Aaron Do}
    \affiliation{Institute for Astronomy, University of Hawai`i at M$\bar{a}$noa, Honolulu, HI 96822, USA}
    \author[0000-0002-6124-1196]{Adam G. Riess}
    \affiliation{Space Telescope Science Institute, Baltimore, MD 21218, USA}
    \affiliation{Department of Physics and Astronomy, Johns Hopkins University, Baltimore, MD 21218, USA}
    \author{Sam M. Ward}
    \affiliation{Institute of Astronomy and Kavli Institute for Cosmology, Madingley Road, Cambridge, CB3 0HA, UK}
    \author[0000-0002-0800-7894]{Arianna Dwomoh}
    \affiliation{Department of Physics, Duke University, Durham, NC 27708, USA}
    \author[0000-0001-6069-1139]{Thomas de Jaeger}
    \affiliation{Institute for Astronomy, University of Hawai`i at M$\bar{a}$noa, Honolulu, HI 96822, USA}
    \affiliation{Laboratoire de Physique Nucleaire et de Hautes-Energies, Barre 12-22 1er etage, 4 place Jussieu, F-75005 Paris, France}
    \author[0000-0001-8738-6011]{Saurabh W. Jha}
    \affiliation{Department of Physics and Astronomy, Rutgers, the State University of New Jersey, Piscataway, NJ 08854, USA}
    \author[0000-0001-9846-4417]{Kaisey S. Mandel}
    \affiliation{Institute of Astronomy and Kavli Institute for Cosmology, Madingley Road, Cambridge, CB3 0HA, UK}
    \author[0000-0002-2361-7201]{Justin D. R. Pierel}
    \affiliation{Space Telescope Science Institute, Baltimore, MD 21218, USA}
    \author[0000-0002-8012-6978]{Brodie Popovic}
    \affiliation{Department of Physics, Duke University, Durham, NC 27708, USA}
    \author[0000-0002-1873-8973]{Benjamin M. Rose}
    \affiliation{Department of Physics, Duke University, Durham, NC 27708, USA}
    \author[0000-0001-5402-4647]{David Rubin}
    \affiliation{Department of Physics and Astronomy, University of Hawai`i at M$\bar{a}$noa, Honolulu, HI 96822, USA}
    \author[0000-0003-4631-1149]{Benjamin J. Shappee}
    \affiliation{Institute for Astronomy, University of Hawai`i at M$\bar{a}$noa, Honolulu, HI 96822, USA}
    \author{Stephen Thorp}
    \affiliation{Institute of Astronomy and Kavli Institute for Cosmology, Madingley Road, Cambridge, CB3 0HA, UK}
    \author[0000-0003-2858-9657]{John L. Tonry}
    \affiliation{Institute for Astronomy, University of Hawai`i at M$\bar{a}$noa, Honolulu, HI 96822, USA}
    \author[0000-0002-9291-1981]{R. Brent Tully}
    \affiliation{Institute for Astronomy, University of Hawai`i at M$\bar{a}$noa, Honolulu, HI 96822, USA}
    \author[0000-0001-8788-1688]{Maria Vincenzi}
    \affiliation{Department of Physics, Duke University, Durham, NC 27708, USA}

\begin{abstract}
While the sample of optical Type Ia Supernova (SN Ia) light curves (LCs) usable for cosmological parameter measurements surpasses 2000, the sample of published, cosmologically viable near-infrared (NIR) SN Ia LCs, which have been shown to be good ``standard candles," is still $\lesssim$ 200.
Here, we present high-quality NIR LCs for 83 SNe Ia ranging from $0.002 < z < 0.09$ as a part of the Dark Energy, H$_0$, and peculiar Velocities using Infrared Light from Supernovae (DEHVILS) survey.
Observations are taken using UKIRT's WFCAM, where the median depth of the images is \Ydepth, \Jdepth, and \Hdepth~mag (Vega) for \textit{Y}, \textit{J}, and \textit{H}-bands, respectively.
The median number of epochs per SN Ia is \totepochs~for all three bands (\textit{YJH}) combined and \Yepochs~for each band individually.
We fit 47 SN Ia LCs that pass strict quality cuts using three LC models, SALT3, SNooPy, and \textsc{BayeSN} and find scatter on the Hubble diagram to be comparable to or better than scatter from optical-only fits in the literature.
Fitting NIR-only LCs, we obtain standard deviations ranging from 0.128--0.135~mag.
Additionally, we present a refined calibration method for transforming 2MASS magnitudes to WFCAM magnitudes using HST CALSPEC stars that results in a 0.03 mag shift in the WFCAM \textit{Y}-band magnitudes.

\end{abstract}
\keywords{Cosmology, cosmology: observations, (stars:) supernovae: general}

\section{Introduction}

\begin{table*}[!hbt]
\caption{Publicly available SN counts with NIR data after cumulative cuts}
\begin{threeparttable}
\begin{tabularx}{\textwidth}{l @{\extracolsep{\fill}} ccccccc}
\toprule
 & Total & DEHVILS (\% of Total) & CSP & CfA & RATIR & SweetSpot & RAISIN \\
\midrule

Before Cuts & 429 & 96 (22\%) & 120 & 94 & 41 & 33 & 45 \\
Type Ia normal\tnote{a} & 368 & 83 (23\%) & 85 & 88 & 38 & 30 & 44 \\
$\geq$ 3 NIR epochs\tnote{b} & 345 & 83 (24\%) & 82 & 86 & 33 & 17 & 44 \\
NIR near peak\tnote{c} & 193 & 71 (37\%) & 48 & 49 & 21 & 4 & 0 \\
$z_\textrm{CMB}>0.01$ & \textbf{171} & \textbf{67 (39\%)} & \textbf{40} & \textbf{39} & \textbf{21} & \textbf{4} & \textbf{0} \\ \bottomrule
\end{tabularx}
\begin{tablenotes}
\item[a] Spectroscopically confirmed as Type Ia and designated as normal (e.g.,~exclusion of Iax, 86G-like, 91T-like, 91bg-like, 06gz-like, 06bt-like, Super-Chandrasekhar, and Ia's interacting with circumstellar matter).
\item[b] Number of unique nights with a NIR observation. \item[c] At least one NIR observation within $\pm 3$ days of fitted NIR maximum.
\end{tablenotes}
\end{threeparttable}
\label{table:surveyoverview}
\end{table*}

Type Ia Supernovae (SNe Ia) can be used to characterize the expansion of our universe \citep[e.g.,][]{Freedman19,Riess22}, and historically, SNe Ia have primarily been studied at optical wavelengths \citep[e.g.,][]{Phillips93,Hamuy96,Riess98,Riess99,Perlmutter99,Jha06,Guy10,Conley11,Betoule14,Jones18,Scolnic18,Brout19,Brout22}.
Analysis in the near-infrared (NIR) has advantages however --- namely that SNe Ia in the NIR are more nearly standard candles as the SN light is less affected by dust and more uniform in luminosity \citep{Meikle00,Krisciunas04,Wood-Vasey08,Mandel09,Folatelli10,Phillips12,Kattner12,Barone-Nugent12,Avelino19,Mandel22}.
But, observing and working with SNe Ia in the NIR is difficult. 
The sky background is much brighter in the NIR than in the optical, making NIR light from SNe more difficult to observe. 
Additionally, galaxies are brighter relative to SNe in the NIR, making light from SNe discovered on top of galaxy light more difficult to extract; SNe Ia are fainter in the NIR; and NIR detectors have historically trailed in both quantity and quality to optical detectors.
Due to these difficulties, fewer SNe Ia have been observed in the NIR, and less work has been done in terms of NIR SN Ia testing, analyzing, and light curve (LC) modeling than in the optical.

NIR SN Ia samples and analyses have been performed by projects such as the Carnegie Supernova Project \citep[CSP;][]{Hamuy06} through CSP-I \citep{Contreras10,Stritzinger11,Krisciunas17} and CSP-II \citep[yet to be publicly available;][]{Phillips19,Hsiao19}, CfA \citep{Wood-Vasey08,Friedman15}, RATIR \citep{Johansson21}, SweetSpot \citep{Weyant18}, RAISIN \citep{RAISIN}, and are ongoing in the VISTA Extragalactic Infrared Legacy Survey (VEILS) and the Supernovae in the InfRAred Avec Hubble (SIRAH) projects.
However, as the sample of LCs in the optical approaches $\sim$2000 \citep{Scolnic22}, the sample of published NIR LCs approaching the typical quality of optical LCs is still $\lesssim$~200.
We present respective SN counts from each of these projects, including this work, in Table~\ref{table:surveyoverview}.
Still, even with relatively few high-quality NIR LCs available,
analysis on SNe Ia in the NIR has resulted in
measurements of the Hubble constant, H$_0$, with a few percent precision \citep[e.g.,][]{Burns18,Dhawan18,RAISIN, Galbany22,DhawanThorp22}.
These measurements show the utility of a sample with largely independent systematic uncertainties.

SN Ia optical brightnesses have been shown to correlate with LC stretch and LC color \citep{Pskovskii77,Phillips93,Tripp98}.
In addition to these empirical correlations accounted for in SN Ia LC analyses, 
for a given best-fit shape and color parameter,
SNe Ia found in galaxies with higher stellar masses are intrinsically brighter in the optical after standardization than those found in less massive host galaxies
\citep{Kelly10,Sullivan10,Lampeitl10}.
This so-called \textit{mass step} has had many theories for its physical explanation, one of which is progenitor physics \citep{Rigault20}, and another explanation is dust, in particular different attenuation dust laws in high and low mass galaxies \citep{BroutScolnic21}.
Since dust absorbs less NIR light, the mass step should starkly decrease or altogether disappear in the NIR.
The findings from tests on this prediction have been varied.
Both \citet{Uddin20} and \citet{Ponder21} claim there is evidence for a mass step in the NIR of $\sim$0.10 $\pm$ 0.04 mag. 
\citet{Johansson21} do not find evidence for a NIR mass step in \textit{J} and \textit{H}-bands ($\sim$0.02 $\pm$ 0.03 mag), while RAISIN \citep{RAISIN} and \citet{Thorp22} present similar findings as \citet{Uddin20} and \citet{Ponder21} with limited significance ($\sim$0.07 $\pm$ 0.04 mag at $10^{10}$ $\textrm{M}_\odot$).

With the Dark Energy, H$_0$, and peculiar Velocities using Infrared Light from Supernovae (DEHVILS) survey, we look to substantially improve the NIR SN Ia data sample, further augment NIR SN Ia understanding, and provide an anchor sample for the upcoming data from the Rubin Observatory \citep{Ivezic19} and Roman Space Telescope \citep{Spergel15,Hounsell18,Rose21}.
One of our main goals is to improve measurements of dark energy parameters (i.e., $w$), H$_0$, and the growth-of-structure parameter $f\sigma_8$ in the nearby universe.
We plan to analyze a potential NIR mass step from DEHVILS in future work.

A complementary project to DEHVILS, Hawaii Supernova Flows (HSF; Do et al.~in prep.), targets far more SNe 
but with fewer epochs and in only one or two filters (primarily \textit{J}-band alone). 
Both \citet{Stanishev18} and \citet{Muller-Bravo22} have demonstrated that constraining the time of peak with optical data and fitting for distance with the NIR (even with few epochs) is viable for cosmological analyses.
HSF will be able to test the claims from \citet{Stanishev18} and \citet{Muller-Bravo22} and measure cosmological parameters with large statistics, while our sample is better suited for building SN standardization models, testing which NIR LC characteristics are best for cosmology, and anchoring Roman across multiple wavelengths.

In Section~\ref{sec:Survey} we describe the framework of the DEHVILS survey images and data, in Section~\ref{sec:Data_Reduction} we illustrate how the data are reduced, and in Section~\ref{sec:Calibration} we describe and validate the photometric calibration.
In Sections~\ref{sec:LCFIT} and~\ref{sec:HD} we fit the SN LCs to different SN models and present the initial analysis of our data.
Discussions and conclusions are in Sections~\ref{sec:Discussion} and~\ref{sec:Conclusions}.

\section{Survey Overview}\label{sec:Survey}
DEHVILS is a follow-up survey with the goal of improving cosmological parameter measurements from SNe Ia in the NIR.
Images from DEHVILS were obtained using the Wide Field Camera\footnote{\url{https://about.ifa.hawaii.edu/ukirt/instruments/wfcam/}.} (WFCAM) mounted on the United Kingdom InfraRed Telescope\footnote{\url{https://about.ifa.hawaii.edu/ukirt/}.} (UKIRT) on Maunakea. UKIRT is a 3.8-meter, reflecting, Cassegrain telescope with an aluminum mirror coating (which was most recently applied in 2010). WFCAM is a NIR wide-field camera with four individual arrays\footnote{Rockwell Hawaii-II HgCdTe 2048$\times$2048 instruments.} each covering 0.05 degrees$^2$ \citep[13.65'$\times$13.65';][]{Casali07}.

A summary of the initial DEHVILS survey data release can be found in Table~\ref{table:surveyoverview}. 
The final counts in Table~\ref{table:surveyoverview} are not statements on how many SNe should be used in a cosmological analysis for a given survey, but instead are intended to give the reader a sense of how many NIR LCs are available with a given set of criteria.
For example, for measuring the Hubble constant with NIR observations, \citet{Galbany22} use even stricter cuts than those given in Table~\ref{table:surveyoverview} utilizing only 37 SNe from CSP and 26 SNe from CfA (with some LCs in both surveys).
\citet{Pierel22} incorporate just 25 CSP SNe and 22 CfA SNe with NIR data in their training sample for a NIR extension of the SALT3 LC fitting model.
There are other sources of published NIR LCs in the literature (Table~\ref{table:surveyoverview} is not an exhaustive list), but the surveys listed are some of the largest and most widely used samples of NIR LCs available.

In total, for this initial data release, DEHVILS observed 96 SNe with 83 SNe spectroscopically confirmed as Type Ia and designated as normal. Of the spectroscopically-confirmed SNe Ia, we have host galaxy spectroscopic redshifts for 77 of them. 
We observe all of our SNe in the NIR \textit{YJH} bands. We reach a median of \totepochs~epochs in all three bands combined for the SNe in our sample,
with \Yepochs~epochs in each of the three bands.
All SNe have at least 3 epochs in \textit{Y} and \textit{H} and at least 2 epochs in \textit{J}.
Information on each individual SN can be found in Table~\ref{tab:SNe} including SN type, coordinates, redshift ($z$), host galaxy (Principal Galaxies Catalog; PGC ID), discovering group, classifying group, and number of epochs.
Of the 12 SNe that have no classification on the Transient Name Server,\footnote{\url{https://www.wis-tns.org/}.} $\sim$1/2 fit reasonable well to a SN Ia template.
Data are available for download at \url{https://github.com/erikpeterson23/DEHVILSDR1}.

\subsection{Target Selection}
As a follow-up survey, DEHVILS
candidates were obtained from many different surveys discovering SNe including the Asteroid Terrestrial-impact Last Alert System \citep[ATLAS;][]{ATLAS,Smith20}, the Zwicky Transient Facility \citep[ZTF;][]{Bellm19} through the Automatic Learning for the Rapid Classification of Events \citep[ALeRCE;][]{Forster21} broker, Supernova and Gravitational Lenses Follow-up \citep[SGLF;][]{Poidevin20jdo,Poidevin20kcr,Shirley20npb,Angel20qic,Perez20tug,Marques20uec,Poidevin21dnm}, Pan-STARRS1 \citep[PS1;][]{Chambers16,Chambers20kbw,Chambers21usd}, GaiaAlerts \citep{Hodgkin21bbz,Hodgkin21ble,Hodgkin21}, the Gravitational-wave Optical Transient Observer \citep[GOTO;][]{Steeghs20oms,Steeghs22}, SIRAH \citep{Jha20pst}, \citet{Itagaki21fxy}, and the Young Supernova Experiment \citep[YSE;][]{Jones21,Jones21zfw}.

LCs of objects recently discovered were analyzed in order to pick targets.
Only candidates which were likely to be SNe Ia were selected as targets.
To decide if a candidate was a likely SN Ia, we inspected the initial rise time; we looked for candidates which rose $\sim$1 mag/day in the optical for at least two consecutive days.
With this strategy, out of all targets we observed at least once, only 21 out of 166 (12.7\%) were spectroscopically confirmed as non-Ia. Classification efforts are further described in Section~\ref{subsec:classification}.

After July 15th, 2020 (after data on about half of our SNe had been taken), we modified our target selection strategy such that all targets needed to be not only likely SNe Ia but also likely to reach 18th mag in the optical (\textit{r} band), again judging from the early-time optical LC. This modification resulted in DEHVILS LCs with slightly later initial phases (relative to optical maximum, median initial phase before this modification was $-5.4$ days and $-4.7$ days after), but the targets were more likely SN Ia and SNe for which we could obtain high signal-to-noise observations more easily. 
For those SNe targeted in SIRAH, introduced above, we intentionally followed regardless of the SN's likelihood of reaching 18th mag.

\subsection{Observing Strategy}
Our preliminary cadence strategy was to aim for 5--6 epochs targeting observations in all \textit{YJH} bands at the phases $-3,0,+3,+8,+13,$ and $+23$ days relative to NIR maximum.
Our goal was to obtain a near 3-day cadence around the NIR-peak and relax to a 5-day cadence and then 10-day cadence post NIR-peak.

Obtaining the exact desired cadence was difficult.
Weather was a factor in stretching out the time between observations as well as the telescope's prioritization requirements.
In the end, we obtained a median cadence of 4.0 days for phases $<+5$ days and a median cadence of 6.6 days for phases $>+5$ days which is greater than but near our preliminary cadence strategy goals.

Because we obtained a poorer cadence than initially intended, we sampled a larger phase range than expected.
Our 5th or 6th epochs (or 7th or 8th epochs when telescope time for our program allowed) reached an estimated phase $\sim+40$ days (we capped observations at 45 days past first epoch), and so we were able to observe and study the secondary maximum present in the NIR SN Ia LC which occurs between rest-frame phase $\sim$20--30 days past NIR peak \citep{Elias81,Kasen06,Mandel09,Folatelli10,Dhawan15,Mandel22}. 
In Section~\ref{sec:LCFIT}, we discuss fitting for a time of maximum in order to obtain LC phases.

We removed targets from the observation queue for a variety of reasons.
First, if a target had not been observed and we estimated it had passed NIR peak ($\sim$2 days before optical peak), the target was removed.
Second, if a target was observed once before or near peak but had not been observed again for over a week, as long as the SN was also past estimated NIR peak, the target was dropped.
And finally, if at any point the target had been spectroscopically classified as non-Ia, the target was dropped.

Of note, when targets in the queue were observed under poor conditions, we view those observations as bonus observations (with albeit poorer quality images). 
These images inflate the total epoch numbers, and the data are used, but we did not count them as fulfilling criteria in our observing strategy in real time.

\setlength\LTleft{0pt}
\setlength\LTright{0pt}
\begin{longtable*}[!hbt]{@{\extracolsep{\fill}} lcccccccr}
\caption{96 DEHVILS SN types, coordinates, heliocentric redshifts, host galaxies, sources, and observation counts}\label{tab:SNe}\\

\toprule
SN & Type & RA & DEC & $z$ & PGC ID & Disc.~Group & Class.~Group & Epochs ($Y$,$J$,$H$) \\
\hline

2020fxa & SN Ia & 15:34:33.040 & +37:32:01.79 & 0.064228 & 2103053 & ALeRCE & SIRAH & 8 (3,2,3) \\
2020jdo & SN Ia & 18:15:43.645 & +58:12:54.94 & 0.072513 & 2575993 & SGLF & SIRAH & 25 (8,9,8) \\
2020jfc & SN Ia & 19:10:07.080 & +40:00:33.66 & 0.027606 & 62844 & ALeRCE & ZTF & 27 (9,9,9) \\
2020jgl & SN Ia & 09:28:58.426 & $-$14:48:19.88 & 0.006765 & 26905 & ATLAS & SIRAH & 15 (6,4,5) \\
2020jht & SN Ia & 11:59:12.293 & +11:30:19.77 & 0.030441 & 1394626 & ZTF & ZTF & 21 (7,7,7) \\
2020jio & -- & 16:58:10.856 & +46:53:47.87 & 0.043982 & -- & ALeRCE & -- & 21 (7,7,7) \\
2020jjf & SN Ia & 15:35:08.262 & +23:56:44.05 & 0.064268 & 1696614 & ZTF & ZTF & 28 (9,10,9) \\
2020jjh & SN Ia & 16:32:17.340 & +50:11:25.94 & 0.047369 & 58466 & ZTF & ZTF & 21 (7,7,7) \\
2020jsa & SN Ia & 14:24:23.974 & +26:41:23.19 & 0.036555 & 51460 & ZTF & ZTF & 25 (8,8,9) \\
2020jwl & SN Ia & 18:18:39.279 & +19:11:23.67 & 0.05798 & 1583731 & ZTF & ZTF & 27 (9,9,9) \\
2020kav & -- & 16:12:28.915 & +45:57:36.23 & 0.093855 & -- & ATLAS & -- & 20 (8,4,8) \\
2020kaz & -- & 11:41:06.900 & +42:25:03.97 & 0.067107 & -- & ATLAS & -- & 16 (6,4,6) \\
2020kbw & SN Ia & 15:25:58.515 & +46:18:44.50 & 0.076315 & 9005037 & PS1 & ZTF & 18 (7,4,7) \\
2020kcr & -- & 15:35:42.822 & +33:09:01.74 & 0.110676 & -- & SGLF & -- & 23 (8,7,8) \\
2020khm & -- & 17:55:36.179 & +17:54:58.38 & 0.093571 & -- & ZTF & -- & 27 (9,8,10) \\
2020kkc & SN Ia & 14:35:23.056 & $-$13:39:14.71 & 0.070025 & 9005038 & ALeRCE & ZTF & 21 (7,7,7) \\
2020kku & SN Ia & 17:32:33.723 & +15:44:09.17 & 0.084701 & 3885592 & ALeRCE & \citetalias{Galbany20kku} & 29 (10,9,10) \\
2020kpx & SN Ia & 15:38:10.150 & +04:46:50.41 & 0.022909 & 55656 & ATLAS & SIRAH & 24 (8,8,8) \\
2020kqv & SN Ia & 20:49:03.001 & $-$31:43:51.86 & 0.074988 & 9005039 & ATLAS & DEHVILS & 22 (9,5,8) \\
2020kru & SN Ia & 19:15:35.903 & +53:20:21.50 & 0.027142 & 2438264 & ATLAS & ZTF & 19 (7,5,7) \\
2020krw & SN Ia & 15:02:45.909 & +58:32:05.31 & 0.075716 & 9005040 & ATLAS & ZTF & 16 (7,3,6) \\
2020kyx & SN Ia & 16:13:45.510 & +22:55:14.38 & 0.031915 & 57542 & ALeRCE & \citetalias{Galbany20kyx} & 21 (7,7,7) \\
2020kzn & SN Ia & 15:38:11.563 & +03:01:40.20 & 0.077754 & 3122970 & ZTF & DEHVILS & 28 (11,6,11) \\
2020lfe & SN Ia & 23:48:18.228 & +33:08:56.28 & 0.036485 & 9005042 & ALeRCE & ZTF & 24 (8,8,8) \\
2020lil & SN Ia & 14:37:59.660 & +09:23:18.74 & 0.030715 & 1364397 & ALeRCE & SIRAH & 21 (7,7,7) \\
2020lsc & SN Ia & 16:14:03.750 & +14:16:57.50 & 0.030294 & 57562 & ZTF & ZTF & 18 (6,6,6) \\
2020lwj & -- & 15:22:50.430 & +16:36:21.78 & 0.061635 & -- & ATLAS & -- & 17 (6,5,6) \\
2020may & SN Ia & 12:23:58.950 & +48:21:18.90 & 0.053362 & 2315646 & ZTF & ZTF & 11 (4,3,4) \\
\multicolumn{9}{r}{(\emph{continued on next page})} \\
2020mbf & SN Ia & 20:23:32.720 & +19:42:39.96 & 0.047349 & 5061218 & ATLAS & ZTF & 27 (9,9,9) \\
2020mby & SN Ia & 17:59:52.479 & +50:29:27.31 & 0.05394 & 2374563 & ALeRCE & ZTF & 18 (6,6,6) \\
2020mdd & SN Ia & 15:51:53.805 & +34:04:24.19 & 0.048779 & 4119139 & ATLAS & ZTF & 19 (6,6,7) \\
2020mnv & SN Ia & 16:40:43.830 & +40:24:52.13 & 0.025885 & 2165441 & ATLAS & \citetalias{Balcon20mnv} & 21 (7,7,7) \\
2020mvp & SN Ia & 14:35:46.540 & +24:43:34.43 & 0.03658 & 52171 & ATLAS & ZTF & 22 (7,7,8) \\
2020naj & SN Ia & 13:59:39.060 & +29:42:36.43 & 0.057356 & 4395882 & ZTF & ZTF & 13 (3,5,5) \\
2020nbo & SN Ia & 15:58:14.150 & +18:06:10.04 & 0.05 & -- & ZTF & ZTF & 15 (5,5,5) \\
2020ndv & SN Ia & 23:09:38.098 & +12:49:48.01 & 0.037416 & 1417025 & ATLAS & ZTF & 25 (8,9,8) \\
2020ned & -- & 00:14:26.470 & $-$24:10:09.48 & 0.026028 & -- & ATLAS & -- & 18 (6,6,6) \\
2020nef & SN Ia & 15:19:59.141 & +20:42:47.35 & 0.040954 & 1634244 & ZTF & \citetalias{Soraisam20nef} & 18 (6,6,6) \\
2020npb & SN Ia & 17:59:58.220 & +44:51:45.61 & 0.038463 & 61249 & SGLF & UCSC & 21 (7,7,7) \\
2020npz & -- & 16:35:42.064 & +07:23:03.10 & -- & -- & ATLAS & -- & 18 (6,6,6) \\
2020nst & -- & 21:49:10.293 & $-$27:58:09.89 & 0.070618 & 190310 & ATLAS & -- & 18 (6,6,6) \\
2020nta & 91bg-like & 16:07:23.350 & +13:53:33.75 & 0.03373 & 57215 & ALeRCE & SIRAH & 17 (6,5,6) \\
2020ocv & SN Ia & 17:46:36.924 & +20:30:17.01 & 0.062942 & 1629162 & ZTF & ZTF & 14 (5,4,5) \\
2020oil & SN Ia & 17:54:25.190 & +15:37:04.84 & 0.040462 & 61077 & ZTF & ZTF & 16 (6,4,6) \\
2020oms & SN Ia & 21:38:10.983 & +06:44:27.73 & 0.065038 & 67077 & GOTO & ZTF & 17 (6,5,6) \\
2020pst & SN Ia & 01:13:12.040 & +02:17:01.82 & 0.046035 & 212683 & SIRAH & SIRAH & 12 (5,3,4) \\
2020qic & SN Ia & 00:15:05.600 & +43:20:35.84 & 0.0485 & 9005050 & SGLF & SIRAH & 12 (4,4,4) \\
2020qne & -- & 01:38:20.210 & $-$28:38:54.24 & 0.030391 & 132876 & ATLAS & -- & 12 (4,4,4) \\
2020rgz & SN Ia & 18:12:01.060 & +51:43:28.49 & 0.028753 & 2398983 & ALeRCE & SIRAH & 20 (7,7,6) \\
2020rlj & SN Ia & 23:01:09.614 & +23:29:14.00 & 0.039761 & 3089722 & ALeRCE & SIRAH & 18 (6,6,6) \\
2020sjo & SN Ia & 04:26:21.950 & $-$10:05:55.72 & 0.031268 & 15090 & ZTF & ZTF & 20 (7,7,6) \\
2020sme & SN Ia & 02:44:20.820 & +14:55:16.68 & 0.045625 & 1470478 & ZTF & ZTF & 22 (9,7,6) \\
2020svo & SN Ia & 02:42:32.033 & $-$00:57:47.30 & 0.038747 & 4136229 & ATLAS & ZTF & 21 (7,7,7) \\
2020swy & SN Ia & 01:23:55.560 & $-$38:00:49.54 & 0.032009 & 5121 & ATLAS & SCAT & 18 (6,6,6) \\
2020szr & SN Ia & 23:09:33.100 & +15:39:33.37 & 0.025281 & 70585 & ATLAS & ZTF & 14 (4,6,4) \\
2020tdy & SN Ia & 16:51:21.918 & +07:51:41.79 & 0.04279 & 59121 & ATLAS & ZTF & 14 (5,4,5) \\
2020tfc & SN Ia & 22:17:00.804 & +30:39:21.07 & 0.039431 & 3959667 & ATLAS & UCSC & 21 (7,8,6) \\
2020tkp & SN Ia & 23:58:10.572 & +22:46:12.45 & 0.034794 & 73077 & ZTF & \citetalias{Pellegrino20tkp} & 18 (6,6,6) \\
2020tpf & SN Ia & 05:25:06.969 & $-$21:14:47.21 & 0.028867 & 3698393 & ATLAS & ZTF & 18 (6,6,6) \\
2020tug & SN Ia & 23:59:27.881 & +17:51:54.77 & 0.045769 & 73166 & SGLF & adH0cc & 18 (6,6,6) \\
2020uea & SN Ia & 02:31:21.169 & +43:27:53.25 & 0.019544 & 9595 & ALeRCE & ZTF & 18 (6,6,6) \\
2020uec & SN Ia & 21:09:48.530 & +15:08:54.89 & 0.027914 & 1476551 & SGLF & ZTF & 18 (6,6,6) \\
2020uek & SN Ia & 00:53:54.913 & $-$31:05:36.80 & 0.031942 & 3169 & ATLAS & SCAT & 18 (6,6,6) \\
2020uen & -- & 05:21:52.846 & $-$27:33:52.56 & 0.032553 & -- & ATLAS & -- & 18 (6,6,6) \\
2020ueq & -- & 04:43:21.510 & $-$33:45:41.26 & -- & -- & ATLAS & -- & 18 (6,6,6) \\
2020unl & SN Ia & 06:35:48.100 & +55:42:39.89 & 0.047154 & 2511708 & ALeRCE & \citetalias{Balcon20unl} & 19 (6,7,6) \\
2020vnr & SN Ia & 00:03:16.553 & +02:15:07.84 & 0.09 & 9005071 & ALeRCE & ZTF & 19 (6,7,6) \\
2020vwv & SN Ia & 06:55:20.540 & $-$36:22:46.92 & 0.03186 & 637506 & ATLAS & SCAT & 12 (4,4,4) \\
2020wcj & SN Ia & 02:52:50.310 & $-$01:13:51.71 & 0.02386 & 213114 & ZTF & ZTF & 17 (6,6,5) \\
2020wgr & SN Ia & 08:23:39.740 & $-$00:10:05.99 & 0.034964 & 23543 & ZTF & ZTF & 16 (5,6,5) \\
2020wtq & SN Ia & 01:13:08.065 & +28:40:37.21 & 0.007 & -- & ZTF & ZTF & 15 (5,5,5) \\
2020yjf & SN Ia & 09:42:35.620 & $-$03:29:24.65 & 0.03818 & 1070612 & ZTF & ZTF & 9 (3,3,3) \\
2020ysl & SN Ia & 08:05:42.080 & $-$09:48:52.27 & 0.036419 & 986045 & ATLAS & SCAT & 13 (4,4,5) \\
2020aczg & SN Ia & 10:01:06.690 & +54:47:22.45 & 0.02506 & 9005030 & ATLAS & SIRAH & 9 (3,3,3) \\
2021J & SN Ia & 12:26:27.011 & +31:13:20.55 & 0.002445 & 40692 & ALeRCE & UCSC & 9 (3,3,3) \\
2021ash & SN Ia & 13:19:42.767 & $-$04:31:45.09 & 0.036002 & 3274255 & ATLAS & ePESSTO+ & 10 (4,2,4) \\
2021aut & SN Ia & 13:27:15.401 & $-$26:46:34.00 & 0.044745 & 764003 & ATLAS & ePESSTO+ & 17 (7,5,5) \\
\multicolumn{9}{r}{(\emph{continued on next page})} \\
2021bbz & SN Ia & 11:45:23.600 & +20:19:28.06 & 0.023313 & 36639 & GaiaAlerts & ePESSTO+ & 14 (6,3,5) \\
2021biz & SN Ia & 12:16:34.755 & +33:31:35.51 & 0.021815 & 39329 & ATLAS & Global SN Proj. & 14 (5,4,5) \\
2021bjy & SN Ia & 08:49:33.430 & +50:52:03.68 & 0.027257 & 24798 & ATLAS & ZTF & 14 (5,4,5) \\
2021bkw & SN Ia & 14:21:13.614 & +20:42:53.99 & 0.018193 & 1634284 & ATLAS & ePESSTO+ & 15 (5,5,5) \\
2021ble & SN Ia & 11:39:03.590 & $-$09:24:37.15 & 0.050527 & 3098267 & GaiaAlerts & ePESSTO+ & 15 (5,5,5) \\
2021dnm & SN Ia & 12:37:24.213 & +23:05:36.74 & 0.046072 & 4334569 & SGLF & ZTF & 17 (6,6,5) \\
2021fof & SN Ia & 14:08:12.794 & $-$08:49:57.76 & 0.04 & -- & ATLAS & SCAT & 21 (7,7,7) \\
2021fxy & SN Ia & 13:13:01.570 & $-$19:30:45.18 & 0.009483 & 45908 & \citetalias{Itagaki21fxy} & SIRAH & 18 (6,6,6) \\
2021ghc & SN Ia & 09:56:32.990 & +00:45:08.24 & 0.046172 & 1174488 & ATLAS & SIRAH & 13 (4,5,4) \\
2021glz & SN Ia & 16:24:52.059 & +45:10:51.36 & 0.069562 & 9005119 & ALeRCE & ZTF & 16 (5,6,5) \\
2021hiz & SN Ia & 12:25:41.670 & +07:13:42.20 & 0.003319 & 40566 & ALeRCE & UCSC & 18 (5,8,5) \\
2021huu & SN Ia & 11:54:26.690 & +55:04:26.33 & 0.046189 & 3475597 & ZTF & ZTF & 19 (7,6,6) \\
2021lug & SN Ia & 17:18:23.136 & $-$00:23:15.48 & 0.04 & -- & ZTF & ZTF & 17 (6,5,6) \\
2021mim & SN Ia & 15:40:21.365 & +07:16:53.50 & 0.038233 & 55759 & ALeRCE & SIRAH & 24 (7,10,7) \\
2021pfs & SN Ia & 14:03:23.580 & $-$06:01:53.90 & 0.00897 & 50084 & ALeRCE & DLT40 & 21 (6,9,6) \\
2021usd & SN Ia & 18:22:03.336 & +36:37:20.39 & 0.02824 & 61781 & PS1 & ZTF & 19 (7,6,6) \\
2021zfq & SN Ia & 19:52:46.620 & +50:32:47.94 & 0.025965 & 3097207 & ZTF & ZTF & 18 (6,6,6) \\
2021zfs & SN Ia & 21:32:29.830 & +05:20:46.72 & 0.01963 & 1281969 & ATLAS & \citetalias{Balcon21zfs} & 18 (6,6,6) \\
2021zfw & SN Ia & 21:31:37.680 & +11:49:58.62 & 0.028837 & 66899 & YSE & UCSC & 18 (6,6,6) \\
\bottomrule

\end{longtable*}

\subsection{Science Images}\label{subsec:science images}

For our program, observation requirements included seeing $< 1.8$ arcseconds and cloud coverage of thin cirrus or better ($<20\%$ attenuation variability). 
Whenever possible, we placed the SN at the center of the detector focal plane.
When guide stars were unavailable in the guiding regions with the SN centered, the pointing was modified so that the observation could be carried out with the SN as close to the center of the detector as possible.

Our observations were taken with both half-integer pixel-offset exposures called microstepping and full-integer pixel-offset exposures called jittering. 
The microstep pattern was incorporated so that better resolution (0.2 arcseconds/pixel) than the native WFCAM pixel scale (0.4 arcseconds/pixel) 
and better sampling of the point-spread functions (PSFs) 
could be achieved.
The microstepping was done by taking separate integrations at precisely half-integer pixel offsets from the original pointing (in our case, 4 integrations with 4.62-arcsecond offsets along each axis).\footnote{\label{fn:wfcam}\url{https://about.ifa.hawaii.edu/ukirt/instruments/wfcam/}.}

The jittering utilized for our images was a 5-point 6.4-arcsecond jitter used to ameliorate effects from hot or bad pixels and other potential flat-field issues. Jittering is done by co-adding images taken at whole-integer pixel offsets (rather than half-integer pixel offsets as in microstepping).
After microstepping and jittering with the target in WFCAM's Camera 3, the same process of microstepping and jittering is done with the target in Camera 2 to minimize effects from deficiencies in either camera and for sky removal.\footnote{See footnote \ref{fn:wfcam}. This process is called \texttt{WFCAM\_FLIP\_SLOW} with the \texttt{JITTER\_FLIP32} recipe by UKIRT.} 
We follow UKIRT's recommendation to use Cameras 2 and 3 since Camera 1 has a quantum efficiency valley, and Camera 4 has a dead column.

\begin{figure*}[!htb]
    \centering
    \includegraphics[width=\textwidth]{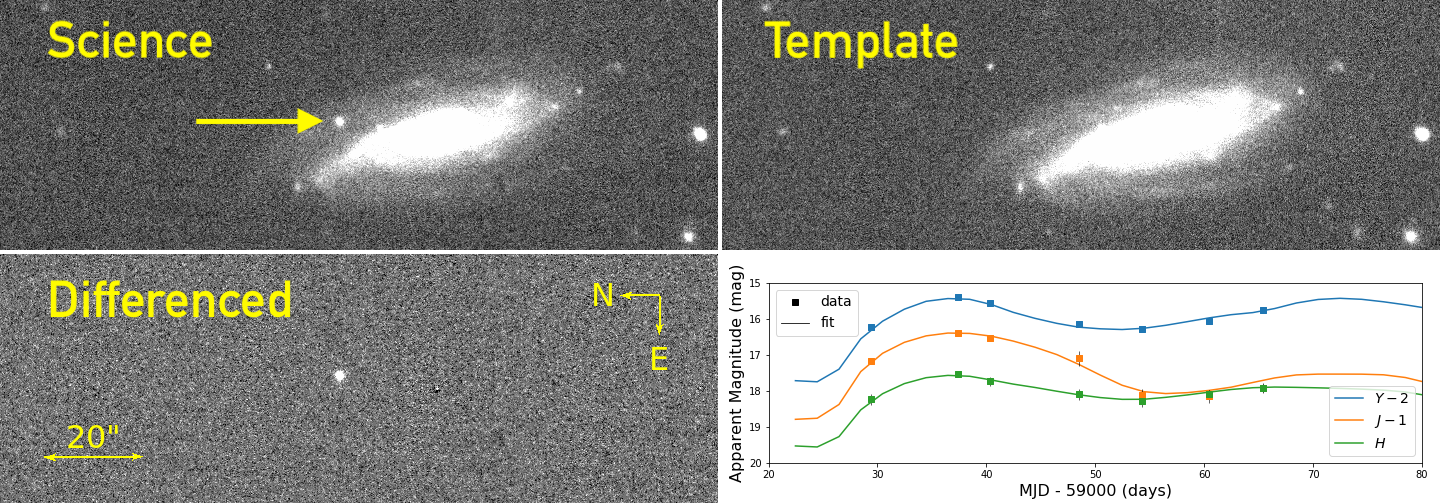}
    \caption{Example images and LC from SN 2020npb. Images span 60''$\times$172'' and are taken in the \textit{Y}-band.
    \textbf{Upper Left}: Science image from MJD 59029.4 (phase $-12.3$ days relative to optical maximum), the first epoch in the LC.
    \textbf{Upper Right}: Template image from MJD 59472.2 (phase $+430.5$ days) taken well over a year past peak brightness.
    \textbf{Lower Left}: Differenced image of the upper two panels with the template image convolved to fit the science image.
    \textbf{Lower Right}: SNooPy \citep{Burns11,Burns14} fit to the LC with magnitude offsets for visualization.
    }
    
    \label{fig:images}
\end{figure*}

Exposure times for all images in all bands were set at 5 seconds. With 2 sets of 5 jitter pointings all microstepping 4 times, each image therefore totaled 200 seconds of exposed time. Across the 3 bands, each target totaled 600 seconds of exposed time. Accounting for time between exposures, filter flushes, slewing, etc.,~total overhead time for one SN epoch was $\sim$1400 seconds (0.4 hours).

By measuring the magnitudes of the faintest detected stars in each image, we reach a median depth of \Ydepth, \Jdepth, and \Hdepth~mag in \textit{Y}, \textit{J}, and \textit{H}, respectively (with standard deviations of \Ydepthstd, \Jdepthstd, and \Hdepthstd~mag).
In the upper-left panel of Fig.~\ref{fig:images}, we present an example science image from SN 2020npb.
The SN itself can be seen offset from the galaxy as indicated by the green arrow a few arcseconds to the upper left of the galaxy.

\subsection{Template Images}\label{subsec:template images}
Template images for the DEHVILS survey were taken in all cases more than a year after our first epoch. Templates were observed and reduced in the exact same way as the corresponding science images except that we doubled the exposure time and required better seeing of $< 1.6$ arcseconds.
The increased exposure time was used to obtain better depth in our template images than in our science images.
The seeing requirement for the template images was modified so that the template image PSFs could be convolved to match the corresponding science image PSFs when performing difference imaging. 
An example template image is provided in the upper-right panel of Fig.~\ref{fig:images} which was taken for SN 2020npb more than a year after peak brightness. When comparing to the corresponding science image to its left, one can see that by eye the SN light is undetectable in the template image.

\subsubsection{Archival Templates}\label{subsubsec:archival templates}
Where available, we used images taken as a part of the UKIRT Infrared Deep Sky Survey \citep[UKIDSS;][]{UKIDSS}\footnote{\url{http://wsa.roe.ac.uk/dr11plus_release.html}.} and the UKIRT Hemisphere Survey \citep[UHS;][]{UHSDR1}\footnote{\url{http://wsa.roe.ac.uk/uhsDR1.html}.} samples as templates rather than obtaining our own. 
The UKIDSS survey ran from 2005--2012,
and the images we obtained from UKIDSS come primarily from the Large Area Survey, but all surveys are queried using the UKIDSSDR11PLUS data release.
Each survey covers approximately 12,700 and 7,000 degrees$^2$ for UHS and UKIDSS respectively.

Images from both UKIDSS and UHS have been published online and were taken on the same instrument (WFCAM) and same telescope (UKIRT) as our sample. Images are obtained using astroquery \citep{astroquery} and coadded and stitched together using SWarp \citep{Bertin02} to construct the templates. 
Since these surveys did not use microstepping, images from these surveys have a larger pixel scale of 0.4 arcseconds/pixel; therefore, when using the archival templates, we resample our corresponding science images to 0.4 arcseconds/pixel using SWarp.

Archival \textit{J}-band templates were obtained from UHS. 
We obtained archival UKIDSS templates for \textit{Y} and \textit{H}-bands but our own templates for \textit{J}-band in the UKIDSS regions. 
When archival UKIDSS templates for \textit{J}-band became accessible for us, we utilized those images as well.
In total, we use 84 archival template images: 14 in \textit{Y}, 55 in \textit{J}, and 15 in \textit{H}. 
Since we implement template subtraction for all science images, the remaining 204 template images required come from DEHVILS as described previously.

\begin{figure*}[!htb]
    \centering
    \includegraphics[width=\textwidth]{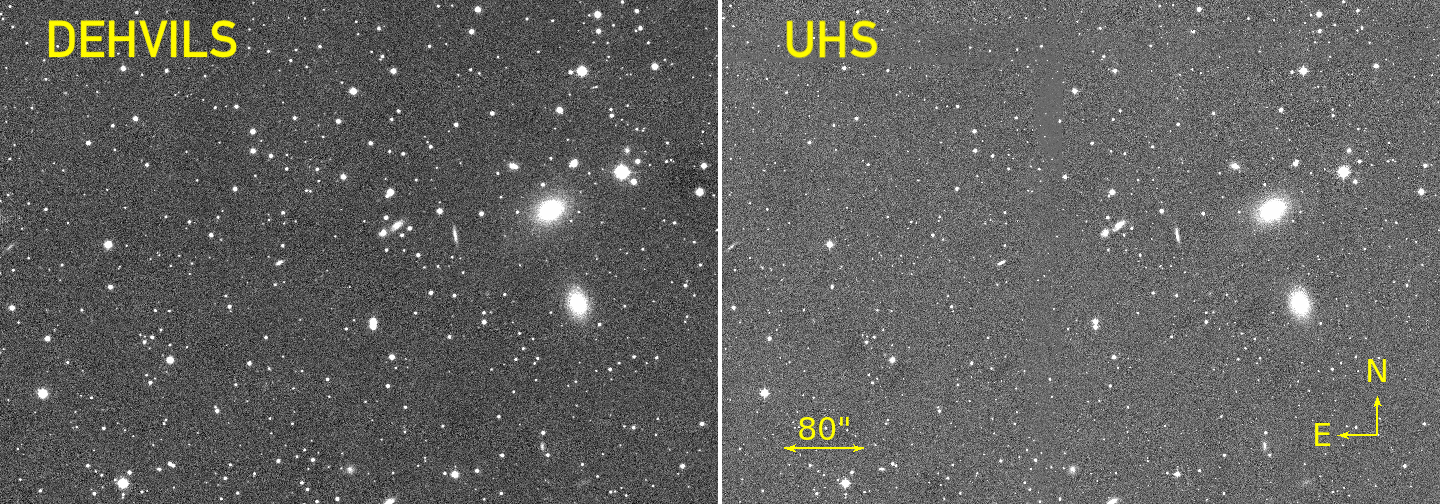}
    \caption{Comparing cutouts from a \textit{J}-band DEHVILS template to an archival \textit{J}-band template image for SN 2020uec.
    \textbf{Left}: Obtained by our program as described in Sections~\ref{subsec:science images}~and~\ref{subsec:template images}.
    \textbf{Right}: Obtained from UHS as described in Section~\ref{subsubsec:archival templates}.}
    
    \label{fig:DEHVILSvUHSDR1}
\end{figure*}

In Fig.~\ref{fig:DEHVILSvUHSDR1}, we compare cutouts from an example DEHVILS template and an archival template from UHS taken at the same location (the SN 2020uec field) and in the same band (\textit{J}).
In many cases, due to the stitching together of multiple survey images, regions of the archival template images that we construct are poor. An effect from this stitching can be seen down the middle of the UHS template.
Additionally, the DEHVILS templates have a smaller pixel scale than the archival templates, so we opt for our own template in the few cases where an archival template was also available.

\subsection{Optical Data}
Optical data for our analysis come from the ATLAS project \citep{ATLAS}.\footnote{\url{https://fallingstar-data.com/forcedphot/}.}
ATLAS samples the accessible night sky from Hawaii on a near two-day cadence in the optical \textit{c} and \textit{o}-bands which span 4200--6500~\AA~and 5600--8200~\AA, respectively. 
The survey reaches a depth of \textit{o} $\sim20$ mag and covers the complete northern sky as well as much of the southern sky down to $\delta\sim-45\degree$. 
ATLAS images are photometrically calibrated using Pan-STARRS \citep{Chambers16}.
LCs from ATLAS are publicly available, and all DEHVILS SN LCs have a corresponding ATLAS LC \citep{Smith20}.
Optical data from ZTF \citep{Bellm19,Forster21} are also available for many of the SNe presented here. An analysis on incorporating data from ZTF will be performed in future work from DEHVILS.

\subsection{Redshifts}
Redshifts come from a variety of sources with host galaxies identified primarily as the nearest neighbor (each SN was inspected individually for assignment).
The University of Hawaii 88-inch telescope (UH88) spectra were analyzed with the same methodology used in Section 6 of \citet{Williams20}, the exception being that we 
avoid the dichroic feature in each SuperNova Integral-Field Spectrograph \citep[SNIFS;][]{Lantz04} spectrum by performing independent cross-correlations for the blue channel ($< 5070$ \AA) and red channel ($> 5170$ \AA). We then visually compare a template spectrum and the two halves of the SNIFS spectrum and use a weighted average as the observed redshift.
The literature sources are from HyperLEDA,\footnote{\url{http://leda.univ-lyon1.fr/}.} which uses multiple published redshift values and a proprietary weighting scheme to estimate the recessional velocity.
Additional redshifts are obtained from Subaru spectra using the same method as used on SNIFS spectra, except for the red/blue separation because the Faint Object Camera and Spectrograph \citep[FOCAS;][]{Kashikawa02} is a slit spectrograph.
68\% of the redshifts come from the literature, 20\% from SNIFS, and 12\% from Subaru.

\begin{figure*}[!htb]
    \centering
    \includegraphics[width=\textwidth]{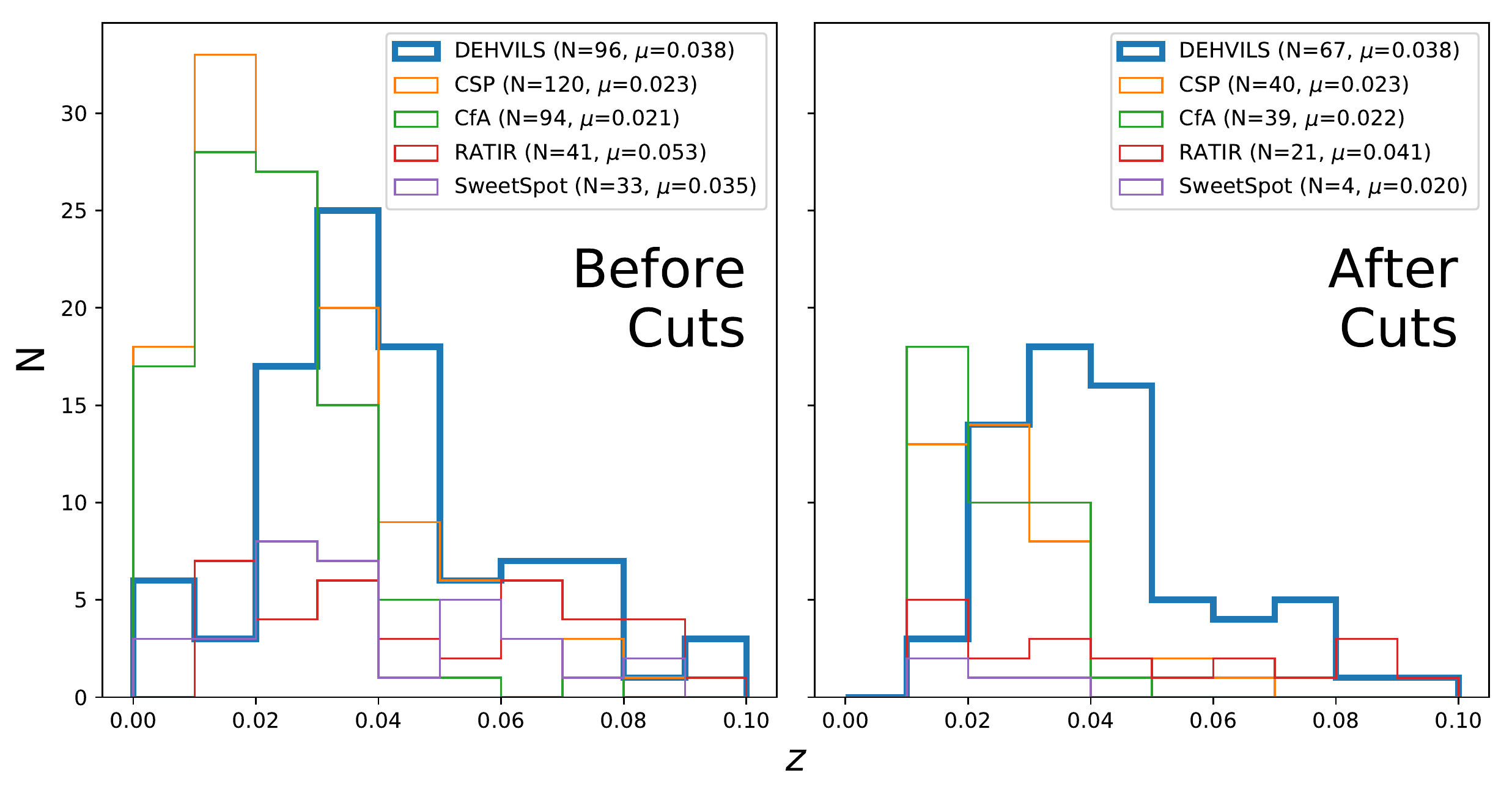}
    \caption{Distribution of redshifts for DEHVILS SNe and SNe from the literature in the first (before cuts; left panel) and final (after cuts; right panel) lines of Table~\ref{table:surveyoverview}.}
    
    \label{fig:z_hist}
\end{figure*}

Following the cuts listed in Table~\ref{table:surveyoverview}, we present the DEHVILS redshift distribution in comparison to other NIR SN Ia redshift distributions from the literature in Fig.~\ref{fig:z_hist}.
RAISIN SNe are not included in the figure since their redshifts are all $>0.1$.
Again, we emphasize that the cuts applied are not definitive statements on which SNe should be used in cosmological analyses (our final sample of SNe on the Hubble diagram is a set of 47 SNe from DEHVILS after cuts, rather than the 67 presented in the right panel of Fig.~\ref{fig:z_hist}), but the figure is instead intended to provide the reader with a sense of how DEHVILS redshifts compare to those from other surveys in the literature.

\subsection{Classification}\label{subsec:classification}
The classifications for DEHVILS SNe come primarily from ZTF.
The second-most common group confirming our targets as Type Ia is SIRAH. 
Other classifying groups given in Table~\ref{tab:SNe} include the University of California Santa Cruz (UCSC) Transients Team, Spectroscopic Classification of Astronomical Transients \citep[SCAT;][]{Tucker22}, the extended-Public ESO Spectroscopic Survey for Transient Objects \citep[ePESSTO+;][]{Smartt15}, \citet{Balcon20mnv,Balcon20unl,Balcon21zfs}, \citet{Galbany20kyx,Galbany20kku}, DEHVILS \citep{Jha20kqv,Jha20kzn}, \citet{Soraisam20nef}, \citet{Pellegrino20tkp}, accurate determination of H$_0$ with core-collapse supernovae \citep[adH0cc;][]{Floers20tug}, the Global SN Project \citep{Burke21biz}, and the Distance Less Than 40 Mpc survey \citep[DLT40;][]{Yang19,Wyatt21pfs}.
The DEHVILS classifications come from spectra obtained using the Robert Stobie Spectrograph (RSS) on the Southern African Large Telescope (SALT).

\section{Data Reduction}\label{sec:Data_Reduction}
Initial reductions for our images from UKIRT are done before we receive the data.
The microstepping and jittering steps are processed and coadded (averaging counts across jitter steps), and the darks and the sky flats for the evening are applied as corrections so that we receive a single semi-processed image with four extensions, one for each camera (Mike Irwin, private communication).

After the initial reductions, we average counts from both WFCAM's Camera 3 and Camera 2 (image acquisition detailed in Section~\ref{subsec:science images}), and we construct mask and noise images given our image characteristics. 
Given that the saturation point for WFCAM is $\sim$60,000--70,000 counts, we set the masking threshold at 40,000 counts as to avoid regions with possible count-rate nonlinearity.
Noise images are computed from the expected poisson noise because read noise for our images is subdominant.

\begin{figure*}[!htb]
    \centering
    \includegraphics[width=\textwidth]{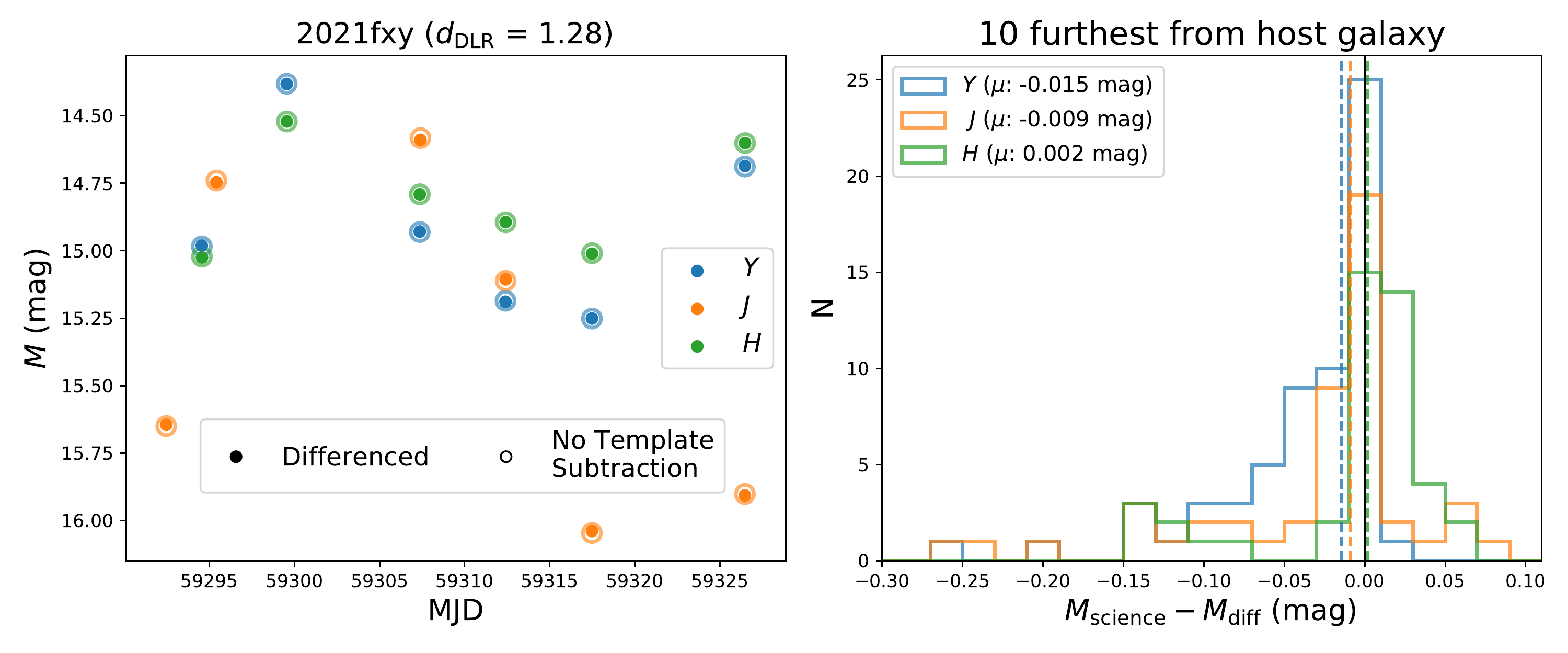}
    \caption{Comparing photometry with and without utilizing a template for galaxy subtraction for SNe relatively far from their host galaxy. 
    \textbf{Left}: Photometry with difference imaging (filled points) and without differencing (unfilled points) for SN 2021fxy which has a large $d_\textrm{DLR}$ (angular distance/host galaxy directional light radius). 
    \textbf{Right}: Residual magnitudes for all epochs comparing with and without differencing for the ten SNe found furthest from their host galaxy. Median values are reported in the legend and plotted as dotted lines. Outliers in this panel are due to clear galaxy contamination viewable by eye for SN 2020jdo and SN 2020swy in \textit{Y} and \textit{J}.}
    
    \label{fig:nosubtraction}
\end{figure*}

Further processing for DEHVILS is based on \textit{photpipe} \citep{Rest05,Rest14}, which is a photometric pipeline that takes SN and template images and outputs calibrated, difference-imaged, forced-photometry LCs.
The \textit{photpipe} reduction pipeline, adapted for the DEHVILS survey here, incorporates the following steps:
\begin{enumerate}
    \item Photometry: PSF Photometry is performed on sources in the science and template images with the \texttt{DoPHOT} software \citep{Schechter93}.
    Stellar detections from this stage are saved and used later in the analysis.
    
    \item Calibration: The 2MASS database is queried for sources within a given WFCAM image footprint. 2MASS \textit{JH} magnitudes are transformed to \textit{YJH} magnitudes in the WFCAM natural system following \citet{Hodgkin09}. 
    Existing \texttt{DoPHOT} photometry for point sources in each image is then compared to the transformed 2MASS photometry in order to calculate a zeropoint. 
    Magnitudes are further refined using Hubble Space Telescope (HST) CALSPEC standard stars. Details on calibration are given in Section~\ref{sec:Calibration}.
    
    \item Difference Imaging:
    After astrometric alignment of the science and template images, difference imaging is carried out using the image subtraction software High Order Transform of PSF ANd Template Subtraction \citep[HOTPANTS;][]{Becker15} which uses the Alard \& Lupton algorithm \citep{Alard98, Alard00}.
    This procedure convolves the template image with three best-fit Gaussian kernels to match the PSF of the science image, scales the template image such that both images have the same zeropoint, and subtracts the template image from the science image.
    
    Alternatively, one could convolve the science images to match the template images; however, our seeing requirement for template images was more restrictive than for the science images and therefore, the average seeing for template images was better than that of the science images. As a result, in all cases we convolved the template image to match the PSF of the science image prior to image subtraction.

    We present an example differenced image in the lower-left panel of Fig.~\ref{fig:images}. 
    
    \item Photometry on the Differenced Image and Final Checks: \texttt{DoPHOT} photometry is performed on the differenced images to determine a weighted-average centroid for each SN. Final forced-photometry LCs including non-detections are produced, again using \texttt{DoPHOT}, for each SN using this centroid.

\end{enumerate}

\begin{figure*}[!htb]
    \centering
    \includegraphics[width=\textwidth]{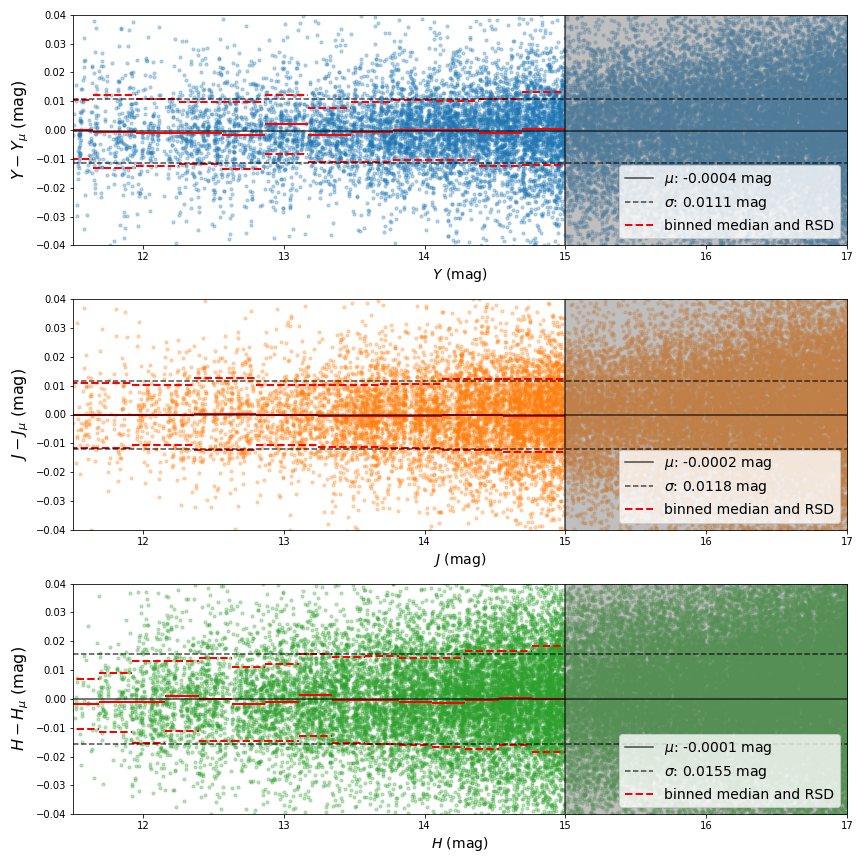}
    \caption{Testing the repeatability of obtaining a given magnitude for a star by taking the distribution of observed magnitudes for that star and comparing it to the mean. Individual magnitudes are compared to the mean magnitude and plotted as a function of magnitude. Binned statistics are plotted in red, and the overall median and RSD of the residuals for all stars brighter than 15 mag are reported in the legend.}
    
    \label{fig:repeatability}
\end{figure*}

\begin{figure*}[!htb]
    \centering
    \includegraphics[width=\textwidth]{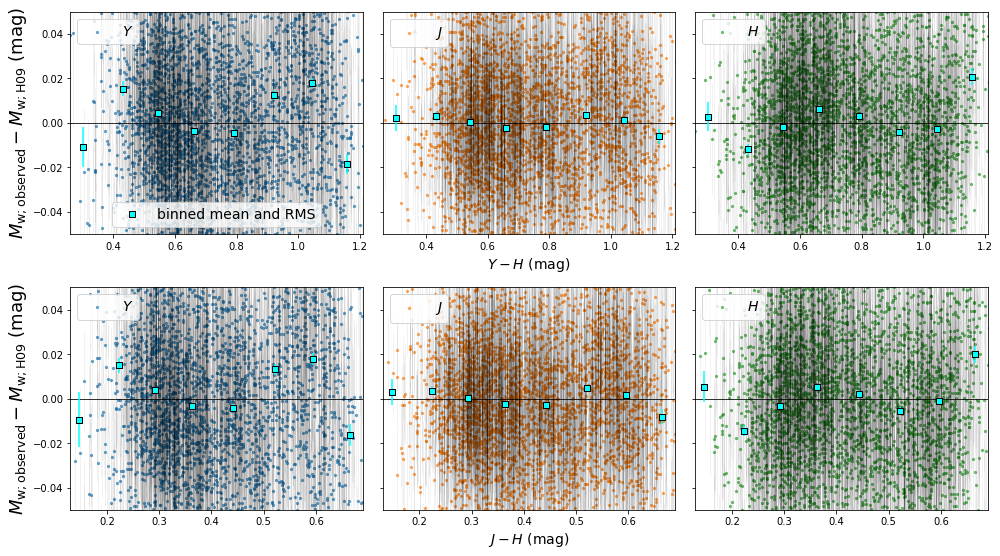}
    \caption{Residual (observed minus transformed from 2MASS) \textit{YJH} WFCAM magnitudes for all stars detected in our sample with respect to color (both \textit{Y}$-$\textit{H} and \textit{J}$-$\textit{H} using transformed magnitudes). Binned means with RMS are also provided.}
    
    \label{fig:calibrations color}
\end{figure*}

\subsection{Validating Image Subtraction}

For the ten SNe furthest from their host galaxy, which were designated using a parameter that uses the angular distance normalized by the host galaxy's directional light radius (DLR; variable dependent on the galaxy's shape and direction to the SN) called $d_\textrm{DLR}$ \citep{Gupta16,Sako18},
photometry could be done on the SN alone and a LC could be constructed without image subtraction. 
The $d_\textrm{DLR}$ values were obtained using the Galaxies HOsting Supernovae and other Transients \citep[GHOST;][]{Gagliano21} program which identifies likely host galaxies for each SN.
These ten SNe all have $d_\textrm{DLR}$ values $>0.98$ meaning their angular separations are all close to if not larger than the host galaxy's DLR, and they provide us with a consistency check for our difference imaging analysis.
The left panel of Fig.~\ref{fig:nosubtraction} depicts how similar LCs (with and without subtraction) are for an example SN designated as far from the galaxy.
For SN 2021fxy, all observations without subtraction are consistent with the values obtained with template subtraction.
In the right panel of Fig.~\ref{fig:nosubtraction}, we present the distribution of residual magnitudes with and without template subtraction observing median residuals of $-0.009$ mag and $0.002$ mag for \textit{J} and \textit{H}, respectively, each near zero.
The robust median absolute standard deviation (RSD, \cite{Hoaglin00}; defined by applying a factor of 1.48 to the median of the absolute values of the deviations of the residuals from the median residual) of these residual magnitudes are 0.022 mag and 0.023 mag. 
We see the largest deviation from zero in \textit{Y}-band at $-0.015$ mag (0.026 mag RSD) indicating photometry on images without subtraction may result in slightly brighter SN magnitudes --- potentially due to galaxy contamination.

\subsection{Repeatability of Stellar Magnitudes}
We check our photometric precision by calculating the repeatability floor of the stars in our sample in Fig.~\ref{fig:repeatability}. We compile stellar detections from all images for 30 of our SNe, calculating the derived magnitude for all stars across all images for each SN. The mean magnitude of each star's distribution of magnitudes is calculated and used to compare distributions of magnitudes across the sample.
Only stars detected at 15th magnitude or brighter are used (the unshaded region of the figure). 
We obtain RSD values of 0.0111, 0.0118, and 0.0155 mag for \textit{Y}, \textit{J}, and \textit{H} respectively, and as a result, we use 0.0150 mag as our repeatability floor. Thus, we include this as an additional error to the calculated SN magnitudes and magnitude errors obtained from the reduction pipeline described in Section~\ref{sec:Data_Reduction}.

\section{Calibration}\label{sec:Calibration}
The general calibration strategy is as follows: 
\begin{enumerate}

\item To determine calibration zeropoints for each image, we use photometry from the Two Micron All Sky Survey \citep[2MASS;][]{Skrutskie06} as it has full sky coverage in \textit{JHK} across our complete survey footprint. This photometry must be transformed to that expected from WFCAM. 

\item To define an absolute calibration, we compute single offsets for each filter by comparing photometry we obtain of HST standard stars in the CALSPEC database \citep{Bohlin14,Bohlin20} to predicted synthetic magnitudes from WFCAM.

\end{enumerate}

\subsection{Calibration Zeropoints for each Image}

To transform the 2MASS photometry to WFCAM magnitudes, we follow \citet{Hodgkin09} (hereafter H09).
The specific transformations for the filters we use are:

\begin{equation}\label{eq:Ycal}
    Y_\textrm{w} = J_2 + 0.500(J_2 - H_2) + 0.080,
\end{equation}

\begin{equation}\label{eq:Jcal}
    J_\textrm{w} = J_2 + 0.065(J_2 - H_2),
\end{equation}

\begin{equation}\label{eq:Hcal}
    H_\textrm{w} = H_2 + 0.070(J_2 - H_2) - 0.030.
\end{equation}

\noindent \textit{YJH} with subscripts w are filter values for WFCAM and are transformed from \textit{JH} 2MASS bands (with subscript~2).
As can be inferred from Eq.~\ref{eq:Ycal}, 2MASS did not have a \textit{Y}-band filter; 2MASS used \textit{JHK} bands.

When we compare our observed stellar magnitudes ($M_\textrm{w; observed}$) to the derived WFCAM magnitudes from the 2MASS catalog ($M_\textrm{w; H09}$; obtained using Eqs.~\ref{eq:Ycal}, \ref{eq:Jcal}, and \ref{eq:Hcal}), we observe the differences in magnitude of the two as a function of color. We present these residuals as a function of 
\textit{Y}$-$\textit{H} and \textit{J}$-$\textit{H} colors
in Fig.~\ref{fig:calibrations color}. We find that across all comparisons when splitting the distributions into bins, the calibrations are reliable down to $2\%$ --- consistent with the findings of H09.
However, we do observe bright stars as $\sim$0.02 mag too bright and magnitude dependent trends in \textit{YJH} of $\sim$8 mmag/mag, $\sim$4 mmag/mag, and $\sim$2 mmag/mag, respectively, but we attribute this to nonlinearity of either the 2MASS or WFCAM detector.

\begin{figure*}[!htb]
    \centering
    \includegraphics[width=\textwidth]{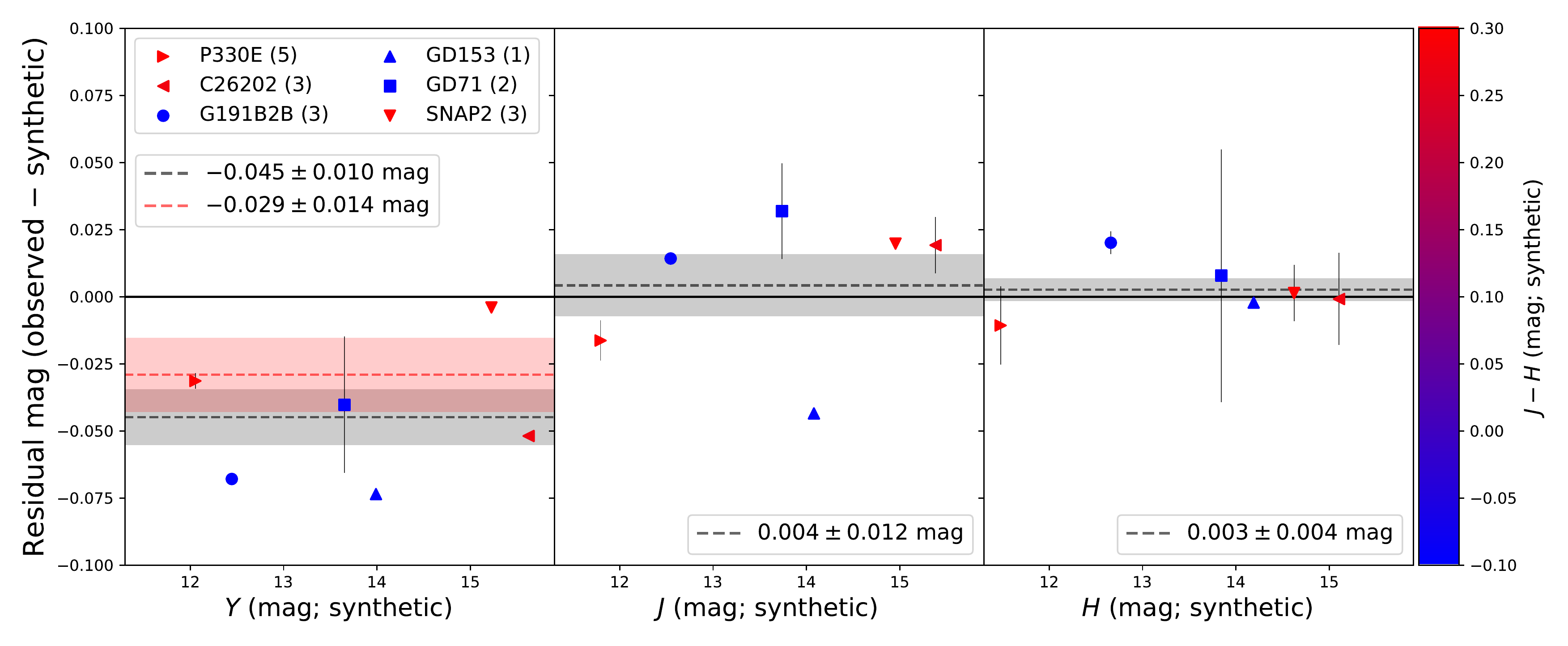}
    \caption{Comparing observed magnitudes to synthetically obtained magnitudes by integrating over the WFCAM bandpasses for six CALSPEC standard stars on the Vega magnitude system. Observed magnitudes are calculated using the obtained flux and the transformation equations from H09 in the zeropoint calculation. For each star, the number of observations from which the mean and standard error on the mean are calculated is reported in the legend. A best-fit offset and relative error is also provided in each panel with an additional fit to the three reddest stars in the \textit{Y}-band panel. \textit{J}$-$\textit{H} colors are depicted as well.}
    
    \label{fig:calspec}
\end{figure*}

\subsection{Absolute Calibration Using HST CALSPEC Stars}
We take observations of HST CALSPEC standard stars in order to further validate and/or improve upon the calibration of our sample.
HST spectra from these CALSPEC standard stars are described in \citet{Bohlin14,Bohlin20} and obtained online.\footnote{\url{https://archive.stsci.edu/hlsps/reference-atlases/cdbs/current_calspec/}.}
Synthetic magnitudes are calculated using those spectra and integrating over the WFCAM \textit{YJH} bandpasses.\footnote{\url{https://about.ifa.hawaii.edu/ukirt/instruments/wfcam/description-of-wfcam/}.}
These synthetic magnitudes are then compared to the observed magnitudes for each of these standard stars in Fig.~\ref{fig:calspec} on the Vega magnitude system.\footnote{CALSPEC stars were compared to the Vega spectrum \texttt{alpha\_lyr\_stis\_005} resulting in offset corrections of 0.624, 0.921, 1.364 mag in $Y$, $J$, and $H$, respectively.}
Most standard stars were observed multiple times, so we provide the mean observed magnitude and standard error on the mean.
Each panel is fit for an offset from zero in gray.
\textit{J}$-$\textit{H} color information is provided on a color scale from blue to red.
We observe that with the current calibration system \textit{J} and \textit{H} seem to be calibrated well (median offsets of $0.004 \pm 0.012$ mag and $0.003 \pm 0.004$ mag, respectively), but \textit{Y} demonstrates a significant offset from zero at $-0.045 \pm 0.010$ mag.

We present an additional fit in red to the three reddest CALSPEC stars in the \textit{Y}-band panel of Fig.~\ref{fig:calspec}.
As can be seen from Fig.~\ref{fig:calibrations color}, these stars, with \textit{J}$-$\textit{H} colors ranging from 0.278--0.331 mag, are the CALSPEC stars most similar (in terms of \textit{J}$-$\textit{H} color) to the stars used to compute zeropoints for DEHVILS.
These three redder stars demonstrate a smaller offset from zero than the full sample of CALSPEC stars in \textit{Y} band, but the offset is still significant at $-0.029 \pm 0.014$ mag.
Given this finding, we add 0.029 mag to all of our SN LC \textit{Y}-band magnitudes and consider the difference between this calculated offset and the offset from the complete CALSPEC sample as a systematic that should be incorporated in cosmological analyses using DEHVILS data.
Additional observations of CALSPEC standard stars will help refine this correction.

\section{Light Curve Model Descriptions and Fits}\label{sec:LCFIT}
We use the \texttt{SNANA} software package \citep{Kessler09, Kessler19} to fit our LCs using both SALT3 \citep{Pierel22} and SNooPy \citep{Burns11,Burns14}.
\texttt{SNANA} is a package that can estimate LC parameters, simulate SN samples, measure distances, and fit for cosmological parameters.
Additional fits are done using \textsc{BayeSN} \citep{Mandel22,Ward22}. For NIR-only fits, the time of maximum is constrained using the optical ATLAS data.

\subsection{SALT3}
We use the SALT3 model \citep{Kenworthy21} for LC fits, but specifically we test the extension to the NIR described in \citet{Pierel22}. The SALT3 model extension is trained on $\sim$1000 LCs with data in the optical bands and 166 LCs with data in the NIR (including data from this sample).
SALT3 is also trained with both optical and NIR spectra to build in K-corrections.
SALT3 fits for a LC stretch, $x_1$, a LC color, $c$, and an amplitude, $x_0$ and uses an equation with flux, $f$, wavelength, $\lambda$, phase, $p$, $M_0$ and $M_1$ as phase-factors on $x_0$ and $x_1$ describing the spectral energy distribution (SED), and \textit{CL} as a color law inputted into the model.
This extension to the SALT3 model \citep{Pierel22}, used for all of our NIR-only, NIR+optical, and optical-only SALT3 fits, expands on the SALT3 model \citep{Kenworthy21} by extending the $M_0$ and $M_1$ wavelength ranges and defining the color law out to 20,000~\AA\ each.

When fitted using the SALT3 model, each LC has a best fit $x_0$, $x_1$, and $c$ value which is then incorporated in the Tripp equation \citep{Tripp98} in order to obtain a distance modulus, $\mu$,

\begin{equation}\label{eq:Tripp_Equation}
    \mu = m_B + {\alpha}x_1 - {\beta}c - \mathcal{M}
\end{equation}

\noindent where $m_B$ is the apparent SN peak magnitude in the $B$ band and directly related to $x_0$ ($m_B = -2.5\log(x_0) + const.$), $\alpha$ and $\beta$ are the regression coefficients that relate $x_1$ and $c$, respectively, to a luminosity correction, and $\mathcal{M}$ is the globally fit absolute SN peak magnitude (for a SN with $c,x_1=0,0$).
When using SALT3 on NIR data, because of the extended framework presented by \citet{Pierel22}, NIR SN LCs are fit in much the same way as in the optical.

\begin{table*}[!hbt]
\caption{SALT3 Fit Parameters and Hubble Residual Scatter}
\begin{threeparttable}
\begin{tabularx}{\textwidth}{l @{\extracolsep{\fill}} cccccccc}
\toprule
Filters & PV Corr. & $x_1$ & $c$ & $\alpha$ & $\beta$ & N fit & RSD & STD \\
 & & & & & & & (mag) & (mag) \\
\midrule
YJH & No & fixed & fixed & 0.000 & 0.000 & 47 & 0.139 $\pm$ 0.026 & 0.172 $\pm$ 0.027 \\
coYJH & No & floated & floated & 0.075 $\pm$ 0.034 & 2.903 $\pm$ 0.223 & 47 & 0.132 $\pm$ 0.025 & 0.175 $\pm$ 0.034 \\
co & No & floated & floated & 0.145 $\pm$ 0.043 & 2.359 $\pm$ 0.249 & 47 & 0.177 $\pm$ 0.029 & 0.221 $\pm$ 0.043 \\
\midrule
YJH & Yes & fixed & fixed & 0.000 & 0.000 & 47 & 0.102 $\pm$ 0.018 & 0.128 $\pm$ 0.015 \\
coYJH & Yes & floated & floated & 0.073 $\pm$ 0.025 & 2.753 $\pm$ 0.161 & 47 & 0.119 $\pm$ 0.019 & 0.128 $\pm$ 0.016 \\
co & Yes & floated & floated & 0.140 $\pm$ 0.034 & 2.349 $\pm$ 0.190 & 47 & 0.141 $\pm$ 0.033 & 0.169 $\pm$ 0.023 \\

\bottomrule
\end{tabularx}
\end{threeparttable}
\label{tab:salt3results}
\end{table*}

\subsection{SNooPy}
SNooPy \citep{Burns11,Burns14} is a model that has been used widely for NIR LC fitting.
SNooPy has been trained on well-calibrated CSP LCs with templates across \textit{uBVgriYJH} bands.
Fitted parameters for SNooPy include a time of maximum, $T_\textrm{max}$, a color-stretch parameter that is a measure of how quickly the SN reaches maximum $B - V$ color, $s_{BV}$, and parameters related to dust; the Milky Way's (MW) color excess $E(B-V)_\textrm{MW}$, the host galaxy's color excess $E(B-V)_\textrm{host}$, and the ratio of total to selective absorption $R_V$.
We use SNooPy's \texttt{EBV\_model2} fixing $s_{BV}=1$, $A_V=0$, and $E(B-V)_\textrm{host}=0$ to fit the NIR-only LCs in \texttt{SNANA} following \citet{RAISIN}.
We also fix the time of maximum to the value determined from the optical data; therefore, the only free parameter in the SNooPy fits is the overall SN amplitude.

\subsection{\textsc{BayeSN}}
Another LC fitter trained on NIR data is \textsc{BayeSN} \citep{Thorp21,Mandel22,Ward22}. \textsc{BayeSN} employs a hierarchical Bayesian approach to modeling SNe Ia, generalizing the previous LC model of \citet{Mandel11}.
\textsc{BayeSN} is a probabilistic time-dependent SED model which leverages both optical and NIR data to infer host galaxy dust parameters and intrinsic spectral components.
Specifically, \textsc{BayeSN} fits the LCs for a distance modulus, $\mu$, the coefficient of the first functional principal component, $\theta$ (correlated strongly with stretch but also includes intrinsic color), and extinction, $A_V$.
For the \textsc{BayeSN} fits, we use the M20 version described in \citet{Mandel22}. Because we are fitting only NIR data, where the sensitivity to dust is minimal, we fix $A_V=0$ and $\theta=-1$ (corresponding to $s_{BV}\approx 1$).

\section{Hubble Diagram}\label{sec:HD}
The distance modulus values for the DEHVILS SNe that are fit using SALT3 and SNooPy are calculated using \texttt{SNANA}.
LC quality cuts for the DEHVILS data are defined as follows: 
SALT3 $|x_1|<3$, uncertainty on SALT3 $x_1 < 1$, uncertainty on the peak MJD $< 2$ days, 
$E(B-V)_\textrm{MW} < 0.2$, and Type Ia LC fit probability (defined by \texttt{SNANA}) $> 0.01$.
Typical LC cuts also include a cut on SALT3 $|c|<0.3$ and uncertainty on SALT3 $c<0.05$, but in the case of NIR-only LC fits, $c$ is often unconstrainable because NIR magnitude is largely degenerate with the $c$ parameter. Thus, we do not include a color constraint in these cases.
In total, 47 out of 83 SNe Ia satisfy all cuts. The LC fit probability cut causes the greatest loss, cutting 40 of the the 46 removed; by visual inspection, we believe the model errors are likely underestimated leading to low fit probabilities, but we leave redevelopment of this model to a future work.

Hubble diagrams are constructed using (i) the $\mu$ values derived from the fitted LC parameters and a modified version of Eq.~\ref{eq:Tripp_Equation}, depending on which corrections are applied (i.e., stretch, color, mass step; could be simply $\mu = m_B - \mathcal{M}$) and (ii) the spectroscopic host redshifts obtained, which are corrected into the Cosmic Microwave Background (CMB) frame.
Where indicated, peculiar velocity (PV) corrections are also applied to the redshifts according to \citet{Peterson22} utilizing the recommended group and coherent-flow corrections.

For measurements of scatter on the Hubble diagram, we calculate the RSD for each of our fitting procedures as the median of the absolute values of the deviations of the Hubble residuals ($\Delta_\mu = \mu-\mu(z)$) from the median residual multiplied by 1.48, 

\begin{equation}
    \textrm{RSD} = 1.48 \times \textrm{median}\big(|\Delta_\mu - \textrm{median}(\Delta_\mu)|\big),
\end{equation}

\noindent where $\mu(z)$ is the predicted distance with a given redshift and a fiducial cosmology.
The sample standard deviation (STD) of the $\mu$ residuals is also provided.
We provide a measurement of the uncertainty in these scatter values using bootstrapping.
We recalculate each scatter value (both RSDs and STDs) choosing the 47 $\mu$ residuals at random, with replacement, 5000 times and take the STD of that sample as a measure of uncertainty for the scatter values given in Table~\ref{tab:salt3results}.

\begin{figure*}[!htb]
    \centering
    \includegraphics[width=\textwidth]{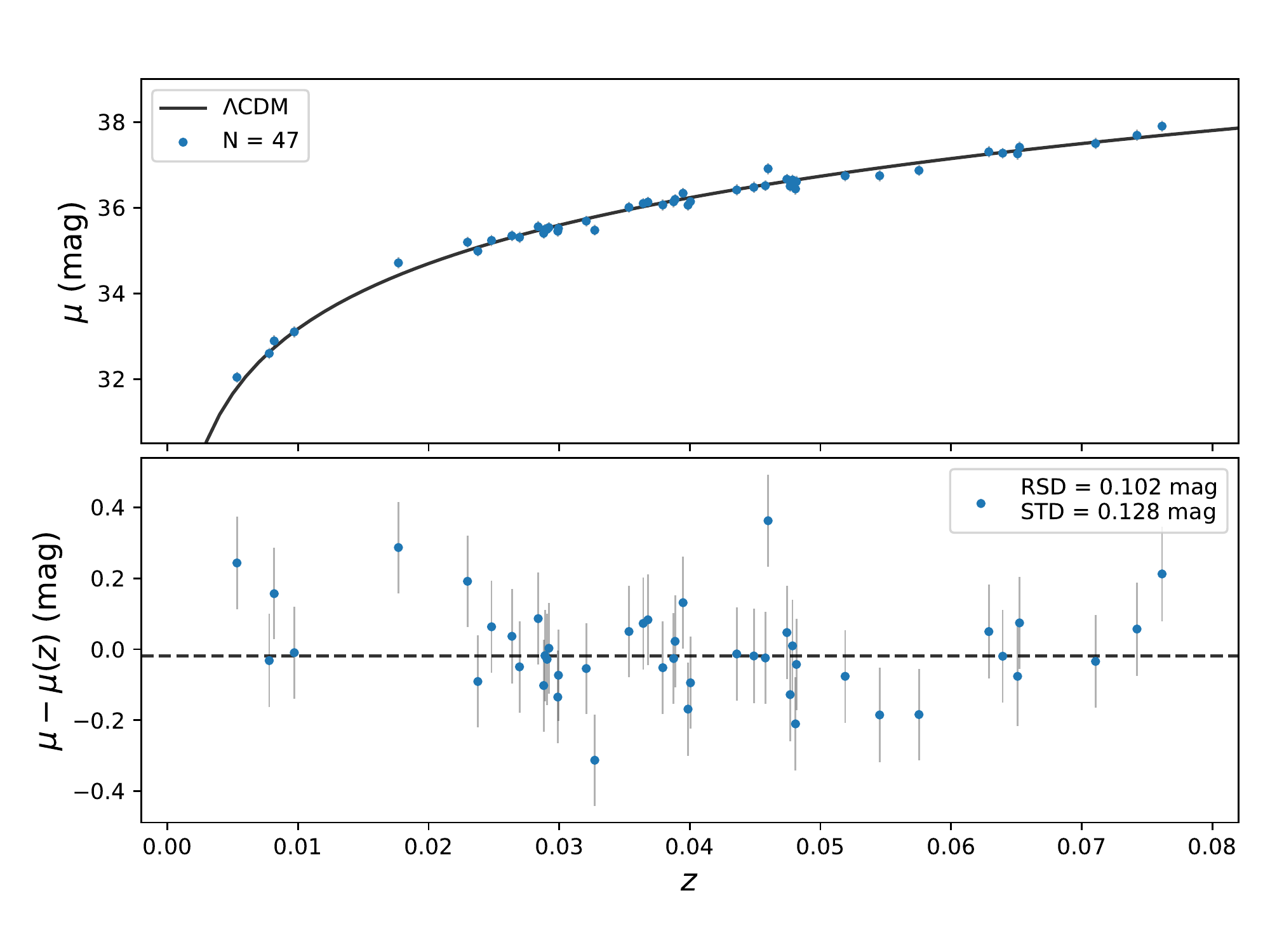}
    \caption{Hubble diagram (top) and Hubble residuals (bottom) for a NIR-only SALT3 fit with the LC parameters $x_1$ and $c$ fixed to zero and PV corrections applied. The dotted line indicates the median Hubble residual of the sample.}
    
    \label{fig:HD_fixx1c_YJH}
\end{figure*}

Studies have claimed that omitting stretch/color corrections in the NIR yields distance measurements with minimal scatter, indicating SNe Ia in the NIR are more nearly standard candles \citep{Dhawan18,Avelino19,Pierel22,Galbany22}.
We test both the omission and inclusion of stretch/color corrections as well as the omission and inclusion of optical data on fits using SALT3. Results from these tests are presented in Table~\ref{tab:salt3results}.
We observe that when comparing NIR-only (\textit{YJH}) SALT3 fits to NIR+optical (\textit{coYJH}) SALT3 fits, scatter is smaller for the NIR-only fits in all cases except for the RSD value without PV corrections.
The inclusion of PV corrections results in improved RSD and STD values for all cases.
When we float $x_1$ while keeping $c$ fixed to zero for NIR-only fits, the scatter increases but the errors on $\alpha$ also increase significantly.
When treating the SNe as perfect standard candles for NIR-only fits (fixing both $x_1=0$ and $c=0$), we observe RSD and STD values of 0.139 mag and 0.172 mag without using PV corrections.
Treating the DEHVILS SNe as standard candles, using only NIR data, and including PV corrections results in the most favorable scatter values (RSD of 0.102 mag and STD of 0.128 mag) in Table~\ref{tab:salt3results}.

\begin{table*}[!hbt]\centering
\caption{LC Model Hubble Residual Scatter Comparison}
\begin{threeparttable}
\begin{tabularx}{\textwidth}{l @{\extracolsep{\fill}} ccccc}
\toprule
Model & Filters & N fit & PV Corr. & RSD & STD \\
 & & & & (mag) & (mag) \\
\hline
SALT3\tnote{a} & YJH & 47 & No & 0.139 $\pm$ 0.026 & 0.172 $\pm$ 0.027 \\
SNooPy & YJH & 47 & No & 0.162 $\pm$ 0.037 & 0.177 $\pm$ 0.031 \\
$\textsc{BayeSN}$ & YJH & 47 & No & 0.115 $\pm$ 0.033 & 0.180 $\pm$ 0.032 \\
\hline
SALT3\tnote{a} & YJH & 47 & Yes & 0.102 $\pm$ 0.018 & 0.128 $\pm$ 0.016 \\
SNooPy & YJH & 47 & Yes & 0.093 $\pm$ 0.026 & 0.135 $\pm$ 0.016 \\
$\textsc{BayeSN}$ & YJH & 47 & Yes & 0.094 $\pm$ 0.025 & 0.131 $\pm$ 0.015 \\
\bottomrule
\end{tabularx}
\begin{tablenotes}
\item[a] fixed $x_1=0$ and fixed $c=0$ from Table \ref{tab:salt3results}.
\end{tablenotes}
\end{threeparttable}
\label{tab:results}
\end{table*}

We present an example Hubble diagram with Hubble residuals with respect to a fiducial cosmology in Fig.~\ref{fig:HD_fixx1c_YJH} which has its results listed in Table~\ref{tab:salt3results}.
The Hubble diagram provided is from the SALT3 fit, fixing both $x_1=0$ and $c=0$ and fitting NIR-only (\textit{YJH}) with PV corrections applied.
When comparing these NIR-only Hubble residuals to the optical-only fitted parameters $x_1$ and $c$, we find no significant correlations ($<2\sigma$ detections of a correlation in each case).
The distance uncertainties obtained for this sample are likely underestimated with the SALT3 fitting process in the NIR (median of 0.030 mag). 
Therefore, the derived intrinsic scatter, $\sigma_\textrm{int}$, to make the reduced $\chi^2$ set to unity for this sample is 0.126 mag, which dominates the errors on $\mu$ in Fig.~\ref{fig:HD_fixx1c_YJH}.

Hubble residual scatter values from all LC fitters are given in Table~\ref{tab:results}. 
Uncertainties on those scatter values are provided and calculated using the same bootstrapping method as described above for Table~\ref{tab:salt3results}.
To ensure a valid comparison between LC models, we use the same SNe in calculating all scatter values,
and LC model fits are NIR-only.
Additionally, we fix similar parameters across each model. For SALT3, we fix both $x_1=0$ and $c=0$; for SNooPy, we fix $A_V=0$ and $s_{BV}=1$; and for \textsc{BayeSN} we fix $A_V=0$ and $\theta=-1$.
All LC models are competitive.
SALT3, SNooPy, and \texttt{BayeSN} exhibit similar results in terms of RSD and STD values.
In terms of RSD, scatter on the Hubble diagram before PV corrections are applied ranges 0.115--0.162 mag and after PV corrections are applied ranges 0.093--0.102 mag. 
In terms of STD, all models exhibit similar values of $\sim$0.175~mag without PV corrections and $\sim$0.130~mag with PV corrections.

\begin{figure*}[!htb]
    \centering
    \includegraphics[width=\textwidth]{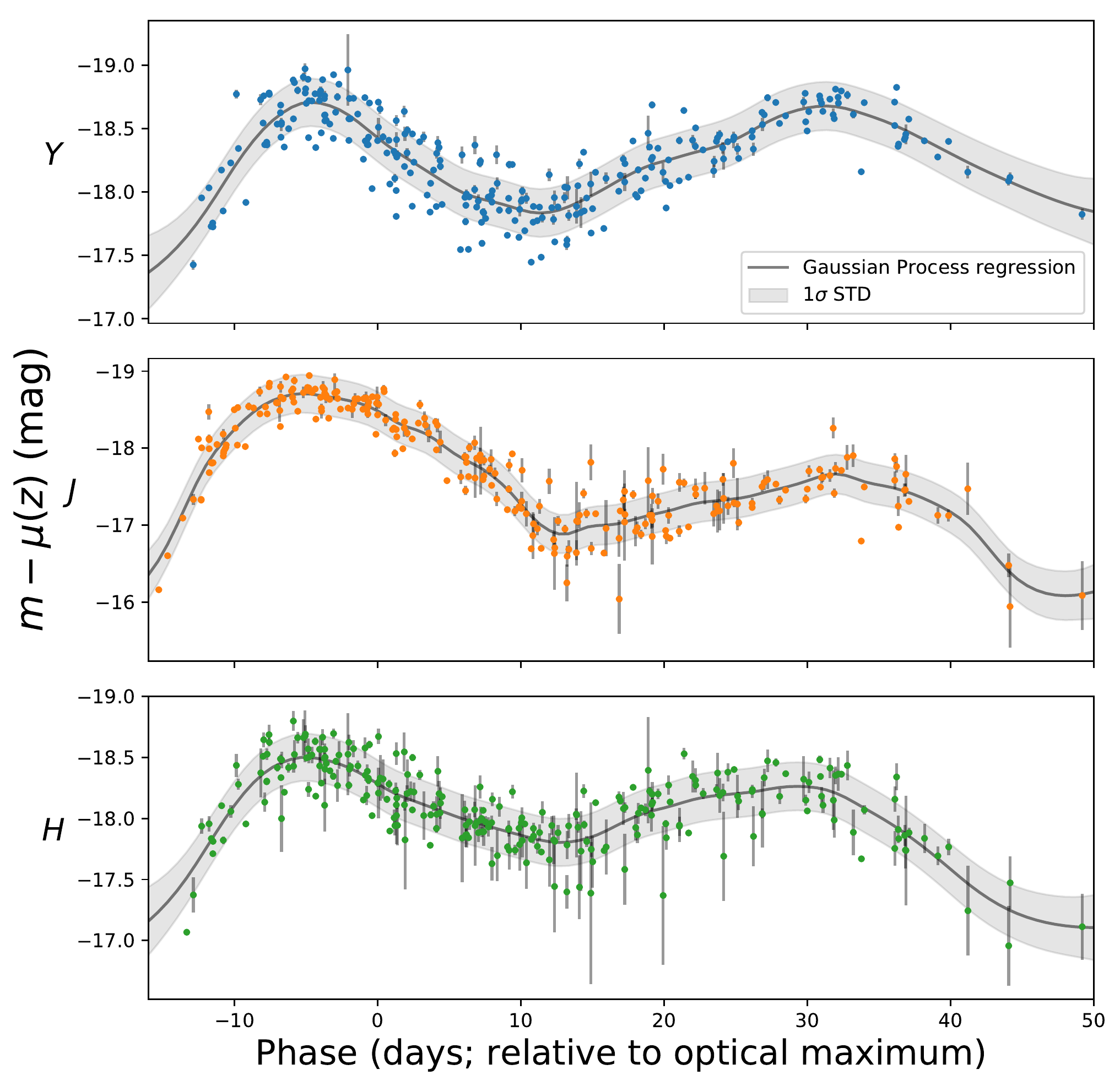}
    \caption{Composite \textit{YJH} LCs for the 47 SNe presented in Fig.~\ref{fig:HD_fixx1c_YJH}. Individual SN LCs are aligned horizontally by phase and vertically by taking the apparent magnitudes and subtracting the redshift-based distance $\mu(z)$ based on a fiducial cosmology. The data are fit using a Gaussian Process regression, and a 1$\sigma$ confidence interval is indicated in gray.}
    
    \label{fig:composite_LC}
\end{figure*}

Using the redshift-based distance $\mu(z)$ based on a fiducial cosmology, we present composite \textit{YJH} LCs in Fig.~\ref{fig:composite_LC} for the 47 SNe passing cuts.
Individual SN LCs are aligned with respect to phase relative to optical maximum and brightness by subtracting $\mu(z)$ values from each SN LC's apparent magnitudes.
A Gaussian Process regression line is provided in Fig.~\ref{fig:composite_LC} along with its STD presented in gray for each filter.
STD values in the DEHVILS composite LCs are 0.189, 0.248, and 0.194 mag at optical maximum for \textit{Y}, \textit{J}, and \textit{H}, respectively.
These results demonstrate how viable treating SNe Ia as standard candles is in the NIR.

\section{Discussion}\label{sec:Discussion}
This initial DEHVILS data release is a sizable contribution to the current publicly available NIR SN Ia LC sample.
In the literature, there are still $\lesssim$ 200 NIR SN Ia LCs with a coverage and cadence similar to DEHVILS. We contribute an additional 83 such LCs to the field.
For the 47 SNe which pass strict SALT3 cuts, we find the best scatter (an RSD value of 0.093 mag) on the Hubble diagram comes from using NIR-only filters, fixing SNooPy $A_V$ and SNooPy $s_{BV}$, and utilizing PV corrections in Table~\ref{tab:results}. 
We note that, for SALT3, the inclusion of optical ATLAS data in the fits results in worse scatter in most cases; in part, this may be due to the fact that the SALT3 model uncertainties do not include off-diagonal covariance to aid in the weighting of data at different wavelengths at this time.
Comparing to the literature, \citet{Stanishev18} compile a sample of $\sim$120 single-band (\textit{J} and \textit{H}) LCs (including data from CSP and CfA) for root-mean-square (RMS) values of $\sim$0.14--0.15 mag.
\citet{Johansson21} report single-band (\textit{J} and \textit{H}) RMS values of $\sim$0.2 mag when combining their data from RATIR with SNe from the literature (e.g., from CSP and CfA) for a total of $\sim$160 SNe and fitting with SNooPy.
\citet{Pierel22} find a NIR-only RMS value of 0.126 mag using SALT3 and a sample of 24 SNe (including a few from this data release).

We aimed for NIR LCs with good sample coverage and achieved a sample average of $\sim$6 epochs in three different bands.
Work has been done previously to ascertain how many epochs and what quality of observations are necessary for good cosmological results with both NIR and optical data \citep{Stanishev18,Muller-Bravo22}.
The findings from these works suggest that NIR epochs need not be plentiful ($\sim$1--2 epochs in a single band) nor outstanding in quality (signal-to-noise $\gtrsim$~8 in \textit{H}) in order to precisely measure cosmological parameters as long as the epochs are near NIR-peak. 
We encourage future works to utilize our data to further test this hypothesis.

In future work, we intend to test a wider range of fitting parameters compared to those presented in both Table~\ref{tab:salt3results} and Table~\ref{tab:results}. For example, we will test single-filter fits and combinations of filters. We plan to examine the effects from using or not using various LC model fitting parameters and from the inclusion of optical data for SNooPy and \textsc{BayeSN} fits. 
Given that our quality cuts are focused on the SALT3 parameters and LC fit metric, we recognize that the scatter results on the selected subsample are likely to favor the SALT3 model, and thus may underestimate the true variation in the NIR LCs.
We also intend to present results from different LC quality-cut criteria (such as cuts focused on results from \textsc{BayeSN} fits) and results from combining our data with LCs in the literature. 
An analysis on a DEHVILS NIR mass step value will also be included in future work.

We also encourage future work to be done on testing the 2MASS to WFCAM filter transformation equations, specifically in the \textit{Y}-band.
Given our HST CALSPEC standard star analysis, we add an additional 0.029 mag to all \textit{Y}-band SN magnitudes as a part of our absolute calibration. 
This systematic could be further tested, for example, with HST observations of the stars in our SN fields.
Analyzing a potential zeropoint systematic due to variation across the field is another goal for future work from DEHVILS.

\section{Conclusions}\label{sec:Conclusions}
In this data release, we present 83 SN Ia LCs. 
We describe both our data reduction pipeline and calibration and confirm that neither introduces significant errors into our analysis.
We validate our HOTPANTS difference imaging process by comparing photometry done on both template-subtracted and unsubtracted images.
We confirm that our calibration is reliable to 2\% by comparing obtained stellar magnitudes to transformed 2MASS catalog stellar magnitudes, and we further refine our calibration by comparing obtained magnitudes to magnitudes synthetically obtained from spectra for HST CALSPEC stars.
We conclude that the \textit{J} and \textit{H} bands are calibrated well, but we find a magnitude offset in \textit{Y} band of 0.029 $\pm$ 0.014 mag indicating our initial derived \textit{Y}-band magnitudes were slightly too bright. We correct for this offset and report our results.

Of the 83 SN Ia LCs, 47 make it through strict SALT3 quality cuts.
We report scatter on the Hubble diagram for these 47 SNe to be $\lesssim$ 0.1 mag using NIR-only fits for the DEHVILS sample which is better than the scatter reported for optical-only Hubble diagrams at $\sim$0.15--0.17 mag \citep{Betoule14,Brout22}. This supports the finding that SNe Ia are excellent standard candles in the NIR.
We find that all the LC models tested here exhibit similar amounts of scatter on the Hubble diagram.

SN Ia cosmology with Roman will require an anchor sample, preferably in the NIR, with comparable precision in calibration and limited systematics.
Continued work and further observations from DEHVILS must be done to achieve this.
The DEHVILS data outlined here have been released to the public at \url{https://github.com/erikpeterson23/DEHVILSDR1}.
Additional data from the DEHVILS survey targeting Cepheids have been published \citep{Konchady22}.

\begin{acknowledgements}
\section*{Acknowledgements}
UKIRT is owned by the University of Hawaii (UH) and operated by the UH Institute for Astronomy. When (some of) the data reported here were obtained, the operations were enabled through the cooperation of the East Asian Observatory.
The authors wish to recognize and acknowledge the very significant cultural role and reverence that the summit of Maunakea has always had within the indigenous Hawaiian community. We are most fortunate to have the opportunity to conduct observations from this mountain.
We would like to thank Mike Irwin for help with data reduction, and Klaus Hodapp, Tom Kerr, Watson Varricatt, and the rest of the UKIRT staff for their support in carrying out this work.

D.S.~is supported by Department of Energy grant DE-SC0010007, the David and Lucile Packard Foundation, the Templeton Foundation and Sloan Foundation.
S.M.W.~and S.T.~were supported by the UK Science and Technology Facilities Council (STFC).
S.W.J.~acknowledges support from the Hubble Space Telescope SIRAH program HST-GO-15889.
K.S.M.~acknowledges funding from the European Research Council under the European Union's Horizon 2020 research and innovation programme (ERC Grant Agreement No.~101002652).
We would also like to thank Maria Acevedo for a careful read of our paper.
This research has made use of NASA's Astrophysics Data System.

\section*{Software}
\texttt{SNANA} \citep{Kessler09}, {astropy} \citep{astropy:2013,astropy:2018},
{matplotlib} \citep{Hunter07},
{numpy} \citep{numpy11}, {PIPPIN} \citep{Pippin}, astroquery \citep{astroquery}, SWarp \citep{Bertin10}.

\section*{Data Availability}
The data are available on GitHub at \url{https://github.com/erikpeterson23/DEHVILSDR1}.
\end{acknowledgements}

\bibliographystyle{mn2e}
\bibliography{main}{}

\begin{thebibliography}{136}
\providecommand{\natexlab}[1]{#1}
\providecommand{\url}[1]{\texttt{#1}}
\providecommand{\urlprefix}{URL }
\providecommand{\eprint}[1][]{\url{#1}}

\bibitem[{{Alard}(2000)}]{Alard00}
{Alard}, C., 2000, \aaps, 144, 363

\bibitem[{{Alard} \& {Lupton}(1998)}]{Alard98}
{Alard}, C., {Lupton}, R.~H., 1998, \apj, 503, 1, 325, \eprint
  arXiv:{astro-ph/9712287}

\bibitem[{{Angel} et~al.(2020){Angel}, {Perez-Fournon} et~al.}]{Angel20qic}
{Angel}, C.~J., {Perez-Fournon}, I., {Poidevin}, F., et~al., 2020, Transient
  Name Server Discovery Report, 2020-2278, 1

\bibitem[{{Astropy Collaboration} et~al.(2013){Astropy Collaboration},
  {Robitaille} et~al.}]{astropy:2013}
{Astropy Collaboration}, {Robitaille}, T.~P., {Tollerud}, E.~J., et~al., 2013,
  \aap, 558, A33, \eprint arXiv:{1307.6212}

\bibitem[{{Avelino} et~al.(2019){Avelino}, {Friedman}, {Mandel}, {Jones},
  {Challis} \& {Kirshner}}]{Avelino19}
{Avelino}, A., {Friedman}, A.~S., {Mandel}, K.~S., {Jones}, D.~O., {Challis},
  P.~J., {Kirshner}, R.~P., 2019, \apj, 887, 1, 106, \eprint arXiv:{1902.03261}

\bibitem[{{Balcon}(2020{\natexlab{a}})}]{Balcon20mnv}
{Balcon}, C., 2020{\natexlab{a}}, Transient Name Server Classification Report,
  2020-2051, 1

\bibitem[{{Balcon}(2020{\natexlab{b}})}]{Balcon20unl}
{Balcon}, C., 2020{\natexlab{b}}, Transient Name Server Classification Report,
  2020-3181, 1

\bibitem[{{Balcon}(2021)}]{Balcon21zfs}
{Balcon}, C., 2021, Transient Name Server Classification Report, 2021-3273, 1

\bibitem[{{Barone-Nugent} et~al.(2012){Barone-Nugent}, {Lidman}
  et~al.}]{Barone-Nugent12}
{Barone-Nugent}, R.~L., {Lidman}, C., {Wyithe}, J.~S.~B., et~al., 2012, \mnras,
  425, 2, 1007, \eprint arXiv:{1204.2308}

\bibitem[{{Becker}(2015)}]{Becker15}
{Becker}, A., 2015, {HOTPANTS: High Order Transform of PSF ANd Template
  Subtraction}, Astrophysics Source Code Library, record ascl:1504.004, \eprint
  ascl:{1504.004}

\bibitem[{{Bellm} et~al.(2019){Bellm}, {Kulkarni} et~al.}]{Bellm19}
{Bellm}, E.~C., {Kulkarni}, S.~R., {Graham}, M.~J., et~al., 2019, \pasp, 131,
  995, 018002, \eprint arXiv:{1902.01932}

\bibitem[{{Bertin}(2010)}]{Bertin10}
{Bertin}, E., 2010, {SWarp: Resampling and Co-adding FITS Images Together},
  Astrophysics Source Code Library, record ascl:1010.068, \eprint
  ascl:{1010.068}

\bibitem[{{Bertin} et~al.(2002){Bertin}, {Mellier}, {Radovich}, {Missonnier},
  {Didelon} \& {Morin}}]{Bertin02}
{Bertin}, E., {Mellier}, Y., {Radovich}, M., {Missonnier}, G., {Didelon}, P.,
  {Morin}, B., 2002, in Astronomical Data Analysis Software and Systems XI,
  edited by {Bohlender}, D.~A., {Durand}, D., {Handley}, T.~H., vol. 281 of
  \emph{Astronomical Society of the Pacific Conference Series}, 228

\bibitem[{{Betoule} et~al.(2014){Betoule}, {Kessler} et~al.}]{Betoule14}
{Betoule}, M., {Kessler}, R., {Guy}, J., et~al., 2014, \aap, 568, A22, \eprint
  arXiv:{1401.4064}

\bibitem[{{Bohlin} et~al.(2014){Bohlin}, {Gordon} \& {Tremblay}}]{Bohlin14}
{Bohlin}, R.~C., {Gordon}, K.~D., {Tremblay}, P.~E., 2014, \pasp, 126, 942,
  711, \eprint arXiv:{1406.1707}

\bibitem[{{Bohlin} et~al.(2020){Bohlin}, {Hubeny} \& {Rauch}}]{Bohlin20}
{Bohlin}, R.~C., {Hubeny}, I., {Rauch}, T., 2020, \aj, 160, 1, 21, \eprint
  arXiv:{2005.10945}

\bibitem[{{Brout} \& {Scolnic}(2021)}]{BroutScolnic21}
{Brout}, D., {Scolnic}, D., 2021, \apj, 909, 1, 26, \eprint arXiv:{2004.10206}

\bibitem[{{Brout} et~al.(2019){Brout}, {Scolnic} et~al.}]{Brout19}
{Brout}, D., {Scolnic}, D., {Kessler}, R., et~al., 2019, \apj, 874, 2, 150,
  \eprint arXiv:{1811.02377}

\bibitem[{{Brout} et~al.(2022){Brout}, {Scolnic} et~al.}]{Brout22}
{Brout}, D., {Scolnic}, D., {Popovic}, B., et~al., 2022, \apj, 938, 2, 110,
  \eprint arXiv:{2202.04077}

\bibitem[{{Burke} et~al.(2021){Burke}, {Gonzalez}, {Hiramatsu}, {Howell},
  {McCully} \& {Pellegrino}}]{Burke21biz}
{Burke}, J., {Gonzalez}, E.~P., {Hiramatsu}, D., {Howell}, D.~A., {McCully},
  C., {Pellegrino}, C., 2021, Transient Name Server Classification Report,
  2021-307, 1

\bibitem[{{Burns} et~al.(2018){Burns}, {Parent} et~al.}]{Burns18}
{Burns}, C.~R., {Parent}, E., {Phillips}, M.~M., et~al., 2018, \apj, 869, 1,
  56, \eprint arXiv:{1809.06381}

\bibitem[{{Burns} et~al.(2011){Burns}, {Stritzinger} et~al.}]{Burns11}
{Burns}, C.~R., {Stritzinger}, M., {Phillips}, M.~M., et~al., 2011, \aj, 141,
  1, 19, \eprint arXiv:{1010.4040}

\bibitem[{{Burns} et~al.(2014){Burns}, {Stritzinger} et~al.}]{Burns14}
{Burns}, C.~R., {Stritzinger}, M., {Phillips}, M.~M., et~al., 2014, \apj, 789,
  1, 32, \eprint arXiv:{1405.3934}

\bibitem[{{Casali} et~al.(2007){Casali}, {Adamson} et~al.}]{Casali07}
{Casali}, M., {Adamson}, A., {Alves de Oliveira}, C., et~al., 2007, \aap, 467,
  2, 777

\bibitem[{{Chambers} et~al.(2020){Chambers}, {Boer} et~al.}]{Chambers20kbw}
{Chambers}, K.~C., {Boer}, T.~D., {Bulger}, J., et~al., 2020, Transient Name
  Server Discovery Report, 2020-1372, 1

\bibitem[{{Chambers} et~al.(2021){Chambers}, {Boer} et~al.}]{Chambers21usd}
{Chambers}, K.~C., {Boer}, T.~D., {Bulger}, J., et~al., 2021, Transient Name
  Server Discovery Report, 2021-2667, 1

\bibitem[{{Chambers} et~al.(2016){Chambers}, {Magnier} et~al.}]{Chambers16}
{Chambers}, K.~C., {Magnier}, E.~A., {Metcalfe}, N., et~al., 2016, arXiv
  e-prints, arXiv:1612.05560, \eprint arXiv:{1612.05560}

\bibitem[{{Conley} et~al.(2011){Conley}, {Guy} et~al.}]{Conley11}
{Conley}, A., {Guy}, J., {Sullivan}, M., et~al., 2011, \apjs, 192, 1, 1,
  \eprint arXiv:{1104.1443}

\bibitem[{{Contreras} et~al.(2010){Contreras}, {Hamuy} et~al.}]{Contreras10}
{Contreras}, C., {Hamuy}, M., {Phillips}, M.~M., et~al., 2010, \aj, 139, 2,
  519, \eprint arXiv:{0910.3330}

\bibitem[{{Dhawan} et~al.(2018){Dhawan}, {Jha} \& {Leibundgut}}]{Dhawan18}
{Dhawan}, S., {Jha}, S.~W., {Leibundgut}, B., 2018, \aap, 609, A72, \eprint
  arXiv:{1707.00715}

\bibitem[{{Dhawan} et~al.(2015){Dhawan}, {Leibundgut}, {Spyromilio} \&
  {Maguire}}]{Dhawan15}
{Dhawan}, S., {Leibundgut}, B., {Spyromilio}, J., {Maguire}, K., 2015, \mnras,
  448, 2, 1345, \eprint arXiv:{1502.00568}

\bibitem[{{Dhawan} et~al.(2022){Dhawan}, {Thorp} et~al.}]{DhawanThorp22}
{Dhawan}, S., {Thorp}, S., {Mandel}, K.~S., et~al., 2022, arXiv e-prints,
  arXiv:2211.07657, \eprint arXiv:{2211.07657}

\bibitem[{{Dye} et~al.(2018){Dye}, {Lawrence} et~al.}]{UHSDR1}
{Dye}, S., {Lawrence}, A., {Read}, M.~A., et~al., 2018, \mnras, 473, 4, 5113,
  \eprint arXiv:{1707.09975}

\bibitem[{{Elias} et~al.(1981){Elias}, {Frogel}, {Hackwell} \&
  {Persson}}]{Elias81}
{Elias}, J.~H., {Frogel}, J.~A., {Hackwell}, J.~A., {Persson}, S.~E., 1981,
  \apjl, 251, L13

\bibitem[{{Floers} et~al.(2020){Floers}, {Taubenberger} et~al.}]{Floers20tug}
{Floers}, A., {Taubenberger}, S., {Vogl}, C., et~al., 2020, Transient Name
  Server Classification Report, 2020-2901, 1

\bibitem[{{Folatelli} et~al.(2010){Folatelli}, {Phillips} et~al.}]{Folatelli10}
{Folatelli}, G., {Phillips}, M.~M., {Burns}, C.~R., et~al., 2010, \aj, 139, 1,
  120, \eprint arXiv:{0910.3317}

\bibitem[{{F{\"o}rster} et~al.(2021){F{\"o}rster}, {Cabrera-Vives}
  et~al.}]{Forster21}
{F{\"o}rster}, F., {Cabrera-Vives}, G., {Castillo-Navarrete}, E., et~al., 2021,
  \aj, 161, 5, 242, \eprint arXiv:{2008.03303}

\bibitem[{{Freedman} et~al.(2019){Freedman}, {Madore} et~al.}]{Freedman19}
{Freedman}, W.~L., {Madore}, B.~F., {Hatt}, D., et~al., 2019, \apj, 882, 1, 34,
  \eprint arXiv:{1907.05922}

\bibitem[{{Friedman} et~al.(2015){Friedman}, {Wood-Vasey} et~al.}]{Friedman15}
{Friedman}, A.~S., {Wood-Vasey}, W.~M., {Marion}, G.~H., et~al., 2015, \apjs,
  220, 1, 9, \eprint arXiv:{1408.0465}

\bibitem[{{Gagliano} et~al.(2021){Gagliano}, {Narayan}, {Engel}, {Carrasco
  Kind} \& {LSST Dark Energy Science Collaboration}}]{Gagliano21}
{Gagliano}, A., {Narayan}, G., {Engel}, A., {Carrasco Kind}, M., {LSST Dark
  Energy Science Collaboration}, 2021, \apj, 908, 2, 170, \eprint
  arXiv:{2008.09630}

\bibitem[{{Galbany} et~al.(2022){Galbany}, {de Jaeger} et~al.}]{Galbany22}
{Galbany}, L., {de Jaeger}, T., {Riess}, A., et~al., 2022, arXiv e-prints,
  arXiv:2209.02546, \eprint arXiv:{2209.02546}

\bibitem[{{Galbany} et~al.(2020{\natexlab{a}}){Galbany}, {Lavers}
  et~al.}]{Galbany20kyx}
{Galbany}, L., {Lavers}, A.~L.~C., {Ashall}, C., et~al., 2020{\natexlab{a}},
  Transient Name Server Classification Report, 2020-1560, 1

\bibitem[{{Galbany} et~al.(2020{\natexlab{b}}){Galbany}, {Lavers},
  {Stritzinger}, {Ashall}, {Morales-Garoffolo} \& {Rosa}}]{Galbany20kku}
{Galbany}, L., {Lavers}, A.~L.~C., {Stritzinger}, M., {Ashall}, C.,
  {Morales-Garoffolo}, A., {Rosa}, N.~E., 2020{\natexlab{b}}, Transient Name
  Server Classification Report, 2020-1420, 1

\bibitem[{{Ginsburg} et~al.(2019){Ginsburg}, {Sip{\H{o}}cz}
  et~al.}]{astroquery}
{Ginsburg}, A., {Sip{\H{o}}cz}, B.~M., {Brasseur}, C.~E., et~al., 2019, \aj,
  157, 3, 98, \eprint arXiv:{1901.04520}

\bibitem[{{Gupta} et~al.(2016){Gupta}, {Kuhlmann} et~al.}]{Gupta16}
{Gupta}, R.~R., {Kuhlmann}, S., {Kovacs}, E., et~al., 2016, \aj, 152, 6, 154,
  \eprint arXiv:{1604.06138}

\bibitem[{{Guy} et~al.(2010){Guy}, {Sullivan} et~al.}]{Guy10}
{Guy}, J., {Sullivan}, M., {Conley}, A., et~al., 2010, \aap, 523, A7, \eprint
  arXiv:{1010.4743}

\bibitem[{{Hamuy} et~al.(2006){Hamuy}, {Folatelli} et~al.}]{Hamuy06}
{Hamuy}, M., {Folatelli}, G., {Morrell}, N.~I., et~al., 2006, \pasp, 118, 839,
  2, \eprint arXiv:{astro-ph/0512039}

\bibitem[{{Hamuy} et~al.(1996){Hamuy}, {Phillips} et~al.}]{Hamuy96}
{Hamuy}, M., {Phillips}, M.~M., {Suntzeff}, N.~B., et~al., 1996, \aj, 112,
  2438, \eprint arXiv:{astro-ph/9609063}

\bibitem[{Hinton \& Brout(2020)}]{Pippin}
Hinton, S., Brout, D., 2020, Journal of Open Source Software, 5, 47, 2122

\bibitem[{Hoaglin et~al.(2000)Hoaglin, Mosteller \& (Editor)}]{Hoaglin00}
Hoaglin, D.~C., Mosteller, F., (Editor), J. W.~T., 2000, Understanding Robust
  and Exploratory Data Analysis, Wiley-Interscience, 1st edn.

\bibitem[{{Hodgkin} et~al.(2021{\natexlab{a}}){Hodgkin}, {Breedt}
  et~al.}]{Hodgkin21bbz}
{Hodgkin}, S.~T., {Breedt}, E., {Delgado}, A., et~al., 2021{\natexlab{a}},
  Transient Name Server Discovery Report, 2021-241, 1

\bibitem[{{Hodgkin} et~al.(2021{\natexlab{b}}){Hodgkin}, {Breedt}
  et~al.}]{Hodgkin21ble}
{Hodgkin}, S.~T., {Breedt}, E., {Delgado}, A., et~al., 2021{\natexlab{b}},
  Transient Name Server Discovery Report, 2021-302, 1

\bibitem[{{Hodgkin} et~al.(2021{\natexlab{c}}){Hodgkin}, {Harrison}
  et~al.}]{Hodgkin21}
{Hodgkin}, S.~T., {Harrison}, D.~L., {Breedt}, E., et~al., 2021{\natexlab{c}},
  \aap, 652, A76, \eprint arXiv:{2106.01394}

\bibitem[{{Hodgkin} et~al.(2009){Hodgkin}, {Irwin}, {Hewett} \&
  {Warren}}]{Hodgkin09}
{Hodgkin}, S.~T., {Irwin}, M.~J., {Hewett}, P.~C., {Warren}, S.~J., 2009,
  \mnras, 394, 2, 675, \eprint arXiv:{0812.3081}

\bibitem[{{Hounsell} et~al.(2018){Hounsell}, {Scolnic} et~al.}]{Hounsell18}
{Hounsell}, R., {Scolnic}, D., {Foley}, R.~J., et~al., 2018, \apj, 867, 1, 23,
  \eprint arXiv:{1702.01747}

\bibitem[{{Hsiao} et~al.(2019){Hsiao}, {Phillips} et~al.}]{Hsiao19}
{Hsiao}, E.~Y., {Phillips}, M.~M., {Marion}, G.~H., et~al., 2019, \pasp, 131,
  995, 014002, \eprint arXiv:{1810.08213}

\bibitem[{Hunter(2007)}]{Hunter07}
Hunter, J.~D., 2007, Computing In Science \& Engineering, 9, 3, 90

\bibitem[{{Itagaki}(2021)}]{Itagaki21fxy}
{Itagaki}, K., 2021, Transient Name Server Discovery Report, 2021-785, 1

\bibitem[{{Ivezi{\'c}} et~al.(2019){Ivezi{\'c}}, {Kahn} et~al.}]{Ivezic19}
{Ivezi{\'c}}, {\v{Z}}., {Kahn}, S.~M., {Tyson}, J.~A., et~al., 2019, \apj, 873,
  2, 111, \eprint arXiv:{0805.2366}

\bibitem[{{Jha} et~al.(2020{\natexlab{a}}){Jha}, {Jones} \& {Do}}]{Jha20kqv}
{Jha}, S., {Jones}, D., {Do}, A., 2020{\natexlab{a}}, Transient Name Server
  Classification Report, 2020-1707, 1

\bibitem[{{Jha} et~al.(2006){Jha}, {Kirshner} et~al.}]{Jha06}
{Jha}, S., {Kirshner}, R.~P., {Challis}, P., et~al., 2006, \aj, 131, 1, 527,
  \eprint arXiv:{astro-ph/0509234}

\bibitem[{{Jha} et~al.(2020{\natexlab{b}}){Jha}, {Dai} et~al.}]{Jha20pst}
{Jha}, S.~W., {Dai}, M., {Perez-Fournon}, I., et~al., 2020{\natexlab{b}},
  Transient Name Server Discovery Report, 2020-2189, 1

\bibitem[{{Jha} et~al.(2020{\natexlab{c}}){Jha}, {Jones} \& {Do}}]{Jha20kzn}
{Jha}, S.~W., {Jones}, D., {Do}, A., 2020{\natexlab{c}}, Transient Name Server
  Classification Report, 2020-1875, 1

\bibitem[{{Johansson} et~al.(2021){Johansson}, {Cenko} et~al.}]{Johansson21}
{Johansson}, J., {Cenko}, S.~B., {Fox}, O.~D., et~al., 2021, \apj, 923, 2, 237,
  \eprint arXiv:{2105.06236}

\bibitem[{{Jones} et~al.(2021{\natexlab{a}}){Jones}, {Foley} et~al.}]{Jones21}
{Jones}, D.~O., {Foley}, R.~J., {Narayan}, G., et~al., 2021{\natexlab{a}},
  \apj, 908, 2, 143, \eprint arXiv:{2010.09724}

\bibitem[{{Jones} et~al.(2021{\natexlab{b}}){Jones}, {French}
  et~al.}]{Jones21zfw}
{Jones}, D.~O., {French}, K.~D., {Agnello}, A., et~al., 2021{\natexlab{b}},
  Transient Name Server Discovery Report, 2021-3267, 1

\bibitem[{{Jones} et~al.(2022){Jones}, {Mandel} et~al.}]{RAISIN}
{Jones}, D.~O., {Mandel}, K.~S., {Kirshner}, R.~P., et~al., 2022, \apj, 933, 2,
  172, \eprint arXiv:{2201.07801}

\bibitem[{{Jones} et~al.(2018){Jones}, {Riess} et~al.}]{Jones18}
{Jones}, D.~O., {Riess}, A.~G., {Scolnic}, D.~M., et~al., 2018, \apj, 867, 2,
  108, \eprint arXiv:{1805.05911}

\bibitem[{{Kasen}(2006)}]{Kasen06}
{Kasen}, D., 2006, \apj, 649, 2, 939, \eprint arXiv:{astro-ph/0606449}

\bibitem[{{Kashikawa} et~al.(2002){Kashikawa}, {Aoki} et~al.}]{Kashikawa02}
{Kashikawa}, N., {Aoki}, K., {Asai}, R., et~al., 2002, \pasj, 54, 6, 819

\bibitem[{{Kattner} et~al.(2012){Kattner}, {Leonard} et~al.}]{Kattner12}
{Kattner}, S., {Leonard}, D.~C., {Burns}, C.~R., et~al., 2012, \pasp, 124, 912,
  114, \eprint arXiv:{1201.2913}

\bibitem[{{Kelly} et~al.(2010){Kelly}, {Hicken}, {Burke}, {Mand el} \&
  {Kirshner}}]{Kelly10}
{Kelly}, P.~L., {Hicken}, M., {Burke}, D.~L., {Mand el}, K.~S., {Kirshner},
  R.~P., 2010, \apj, 715, 2, 743, \eprint arXiv:{0912.0929}

\bibitem[{{Kenworthy} et~al.(2021){Kenworthy}, {Jones} et~al.}]{Kenworthy21}
{Kenworthy}, W.~D., {Jones}, D.~O., {Dai}, M., et~al., 2021, \apj, 923, 2, 265,
  \eprint arXiv:{2104.07795}

\bibitem[{{Kessler} et~al.(2009){Kessler}, {Becker} et~al.}]{Kessler09}
{Kessler}, R., {Becker}, A.~C., {Cinabro}, D., et~al., 2009, \apjs, 185, 1, 32,
  \eprint arXiv:{0908.4274}

\bibitem[{{Kessler} et~al.(2019){Kessler}, {Brout} et~al.}]{Kessler19}
{Kessler}, R., {Brout}, D., {D'Andrea}, C.~B., et~al., 2019, \mnras, 485, 1,
  1171, \eprint arXiv:{1811.02379}

\bibitem[{{Konchady} et~al.(2022){Konchady}, {Oelkers} et~al.}]{Konchady22}
{Konchady}, T., {Oelkers}, R.~J., {Jones}, D.~O., et~al., 2022, \apjs, 258, 2,
  24, \eprint arXiv:{2112.04597}

\bibitem[{{Krisciunas} et~al.(2017){Krisciunas}, {Contreras}
  et~al.}]{Krisciunas17}
{Krisciunas}, K., {Contreras}, C., {Burns}, C.~R., et~al., 2017, \aj, 154, 5,
  211, \eprint arXiv:{1709.05146}

\bibitem[{{Krisciunas} et~al.(2004){Krisciunas}, {Phillips} \&
  {Suntzeff}}]{Krisciunas04}
{Krisciunas}, K., {Phillips}, M.~M., {Suntzeff}, N.~B., 2004, \apjl, 602, 2,
  L81, \eprint arXiv:{astro-ph/0312626}

\bibitem[{{Lampeitl} et~al.(2010){Lampeitl}, {Smith} et~al.}]{Lampeitl10}
{Lampeitl}, H., {Smith}, M., {Nichol}, R.~C., et~al., 2010, \apj, 722, 1, 566,
  \eprint arXiv:{1005.4687}

\bibitem[{{Lantz} et~al.(2004){Lantz}, {Aldering} et~al.}]{Lantz04}
{Lantz}, B., {Aldering}, G., {Antilogus}, P., et~al., 2004, in Optical Design
  and Engineering, edited by {Mazuray}, L., {Rogers}, P.~J., {Wartmann}, R.,
  vol. 5249 of \emph{Society of Photo-Optical Instrumentation Engineers (SPIE)
  Conference Series}, 146--155

\bibitem[{{Lawrence} et~al.(2007){Lawrence}, {Warren} et~al.}]{UKIDSS}
{Lawrence}, A., {Warren}, S.~J., {Almaini}, O., et~al., 2007, \mnras, 379, 4,
  1599, \eprint arXiv:{astro-ph/0604426}

\bibitem[{{Mandel} et~al.(2011){Mandel}, {Narayan} \& {Kirshner}}]{Mandel11}
{Mandel}, K.~S., {Narayan}, G., {Kirshner}, R.~P., 2011, \apj, 731, 2, 120,
  \eprint arXiv:{1011.5910}

\bibitem[{{Mandel} et~al.(2022){Mandel}, {Thorp}, {Narayan}, {Friedman} \&
  {Avelino}}]{Mandel22}
{Mandel}, K.~S., {Thorp}, S., {Narayan}, G., {Friedman}, A.~S., {Avelino}, A.,
  2022, \mnras, 510, 3, 3939, \eprint arXiv:{2008.07538}

\bibitem[{{Mandel} et~al.(2009){Mandel}, {Wood-Vasey}, {Friedman} \&
  {Kirshner}}]{Mandel09}
{Mandel}, K.~S., {Wood-Vasey}, W.~M., {Friedman}, A.~S., {Kirshner}, R.~P.,
  2009, \apj, 704, 1, 629, \eprint arXiv:{0908.0536}

\bibitem[{{Marques-Chaves} et~al.(2020){Marques-Chaves}, {Perez-Fournon}
  et~al.}]{Marques20uec}
{Marques-Chaves}, R., {Perez-Fournon}, I., {Angel}, C.~J., et~al., 2020,
  Transient Name Server Discovery Report, 2020-2935, 1

\bibitem[{{Meikle}(2000)}]{Meikle00}
{Meikle}, W.~P.~S., 2000, \mnras, 314, 4, 782, \eprint arXiv:{astro-ph/9912123}

\bibitem[{{M{\"u}ller-Bravo} et~al.(2022){M{\"u}ller-Bravo}, {Galbany}
  et~al.}]{Muller-Bravo22}
{M{\"u}ller-Bravo}, T.~E., {Galbany}, L., {Karamehmetoglu}, E., et~al., 2022,
  \aap, 665, A123, \eprint arXiv:{2207.04780}

\bibitem[{{Pellegrino} et~al.(2020){Pellegrino}, {Jha}
  et~al.}]{Pellegrino20tkp}
{Pellegrino}, C., {Jha}, S.~W., {Burke}, J., et~al., 2020, Transient Name
  Server Classification Report, 2020-2883, 1

\bibitem[{{Perez-Fournon} et~al.(2020){Perez-Fournon}, {Angel}
  et~al.}]{Perez20tug}
{Perez-Fournon}, I., {Angel}, C.~J., {Poidevin}, F., et~al., 2020, Transient
  Name Server Discovery Report, 2020-2870, 1

\bibitem[{{Perlmutter} et~al.(1999){Perlmutter}, {Aldering}
  et~al.}]{Perlmutter99}
{Perlmutter}, S., {Aldering}, G., {Goldhaber}, G., et~al., 1999, \apj, 517, 2,
  565, \eprint arXiv:{astro-ph/9812133}

\bibitem[{{Peterson} et~al.(2022){Peterson}, {Kenworthy} et~al.}]{Peterson22}
{Peterson}, E.~R., {Kenworthy}, W.~D., {Scolnic}, D., et~al., 2022, \apj, 938,
  2, 112, \eprint arXiv:{2110.03487}

\bibitem[{{Phillips}(1993)}]{Phillips93}
{Phillips}, M.~M., 1993, \apjl, 413, L105

\bibitem[{{Phillips}(2012)}]{Phillips12}
{Phillips}, M.~M., 2012, \pasa, 29, 4, 434, \eprint arXiv:{1111.4463}

\bibitem[{{Phillips} et~al.(2019){Phillips}, {Contreras} et~al.}]{Phillips19}
{Phillips}, M.~M., {Contreras}, C., {Hsiao}, E.~Y., et~al., 2019, \pasp, 131,
  995, 014001, \eprint arXiv:{1810.09252}

\bibitem[{{Pierel} et~al.(2022){Pierel}, {Jones} et~al.}]{Pierel22}
{Pierel}, J.~D.~R., {Jones}, D.~O., {Kenworthy}, W.~D., et~al., 2022, \apj,
  939, 1, 11, \eprint arXiv:{2209.05594}

\bibitem[{{Poidevin} et~al.(2020{\natexlab{a}}){Poidevin}, {Perez-Fournon}
  et~al.}]{Poidevin20jdo}
{Poidevin}, F., {Perez-Fournon}, I., {Angel}, C.~J., et~al.,
  2020{\natexlab{a}}, Transient Name Server Discovery Report, 2020-1226, 1

\bibitem[{{Poidevin} et~al.(2020{\natexlab{b}}){Poidevin}, {Perez-Fournon}
  et~al.}]{Poidevin20kcr}
{Poidevin}, F., {Perez-Fournon}, I., {Angel}, C.~J., et~al.,
  2020{\natexlab{b}}, Transient Name Server Discovery Report, 2020-1374, 1

\bibitem[{{Poidevin} et~al.(2021){Poidevin}, {Perez-Fournon}
  et~al.}]{Poidevin21dnm}
{Poidevin}, F., {Perez-Fournon}, I., {Angel}, C.~J., et~al., 2021, Transient
  Name Server Discovery Report, 2021-556, 1

\bibitem[{{Ponder} et~al.(2021){Ponder}, {Wood-Vasey} et~al.}]{Ponder21}
{Ponder}, K.~A., {Wood-Vasey}, W.~M., {Weyant}, A., et~al., 2021, \apj, 923, 2,
  197, \eprint arXiv:{2006.13803}

\bibitem[{{Price-Whelan} et~al.(2018){Price-Whelan}, {Sip{\H{o}}cz}
  et~al.}]{astropy:2018}
{Price-Whelan}, A.~M., {Sip{\H{o}}cz}, B.~M., {G{\"u}nther}, H.~M., et~al.,
  2018, \aj, 156, 123

\bibitem[{{Pskovskii}(1977)}]{Pskovskii77}
{Pskovskii}, I.~P., 1977, \sovast, 21, 675

\bibitem[{{Rest} et~al.(2014){Rest}, {Scolnic} et~al.}]{Rest14}
{Rest}, A., {Scolnic}, D., {Foley}, R.~J., et~al., 2014, \apj, 795, 1, 44,
  \eprint arXiv:{1310.3828}

\bibitem[{{Rest} et~al.(2005){Rest}, {Stubbs} et~al.}]{Rest05}
{Rest}, A., {Stubbs}, C., {Becker}, A.~C., et~al., 2005, \apj, 634, 2, 1103,
  \eprint arXiv:{astro-ph/0509240}

\bibitem[{{Riess} et~al.(1998){Riess}, {Filippenko} et~al.}]{Riess98}
{Riess}, A.~G., {Filippenko}, A.~V., {Challis}, P., et~al., 1998, \aj, 116, 3,
  1009, \eprint arXiv:{astro-ph/9805201}

\bibitem[{{Riess} et~al.(1999){Riess}, {Kirshner} et~al.}]{Riess99}
{Riess}, A.~G., {Kirshner}, R.~P., {Schmidt}, B.~P., et~al., 1999, \aj, 117, 2,
  707, \eprint arXiv:{astro-ph/9810291}

\bibitem[{{Riess} et~al.(2022){Riess}, {Yuan} et~al.}]{Riess22}
{Riess}, A.~G., {Yuan}, W., {Macri}, L.~M., et~al., 2022, \apjl, 934, 1, L7,
  \eprint arXiv:{2112.04510}

\bibitem[{{Rigault} et~al.(2020){Rigault}, {Brinnel} et~al.}]{Rigault20}
{Rigault}, M., {Brinnel}, V., {Aldering}, G., et~al., 2020, \aap, 644, A176,
  \eprint arXiv:{1806.03849}

\bibitem[{{Rose} et~al.(2021){Rose}, {Baltay} et~al.}]{Rose21}
{Rose}, B.~M., {Baltay}, C., {Hounsell}, R., et~al., 2021, arXiv e-prints,
  arXiv:2111.03081, \eprint arXiv:{2111.03081}

\bibitem[{{Sako} et~al.(2018){Sako}, {Bassett} et~al.}]{Sako18}
{Sako}, M., {Bassett}, B., {Becker}, A.~C., et~al., 2018, \pasp, 130, 988,
  064002, \eprint arXiv:{1401.3317}

\bibitem[{{Schechter} et~al.(1993){Schechter}, {Mateo} \& {Saha}}]{Schechter93}
{Schechter}, P.~L., {Mateo}, M., {Saha}, A., 1993, \pasp, 105, 1342

\bibitem[{{Scolnic} et~al.(2022){Scolnic}, {Brout} et~al.}]{Scolnic22}
{Scolnic}, D., {Brout}, D., {Carr}, A., et~al., 2022, \apj, 938, 2, 113,
  \eprint arXiv:{2112.03863}

\bibitem[{{Scolnic} et~al.(2018){Scolnic}, {Jones} et~al.}]{Scolnic18}
{Scolnic}, D.~M., {Jones}, D.~O., {Rest}, A., et~al., 2018, \apj, 859, 2, 101,
  \eprint arXiv:{1710.00845}

\bibitem[{{Shirley} et~al.(2020){Shirley}, {Perez-Fournon}
  et~al.}]{Shirley20npb}
{Shirley}, R., {Perez-Fournon}, I., {Angel}, C.~J., et~al., 2020, Transient
  Name Server Discovery Report, 2020-1957, 1

\bibitem[{{Skrutskie} et~al.(2006){Skrutskie}, {Cutri} et~al.}]{Skrutskie06}
{Skrutskie}, M.~F., {Cutri}, R.~M., {Stiening}, R., et~al., 2006, \aj, 131, 2,
  1163

\bibitem[{{Smartt} et~al.(2015){Smartt}, {Valenti} et~al.}]{Smartt15}
{Smartt}, S.~J., {Valenti}, S., {Fraser}, M., et~al., 2015, \aap, 579, A40,
  \eprint arXiv:{1411.0299}

\bibitem[{{Smith} et~al.(2020){Smith}, {Smartt} et~al.}]{Smith20}
{Smith}, K.~W., {Smartt}, S.~J., {Young}, D.~R., et~al., 2020, \pasp, 132,
  1014, 085002, \eprint arXiv:{2003.09052}

\bibitem[{{Soraisam} et~al.(2020){Soraisam}, {Lee} et~al.}]{Soraisam20nef}
{Soraisam}, M., {Lee}, C., {Narayan}, G., et~al., 2020, Transient Name Server
  Classification Report, 2020-2147, 1

\bibitem[{{Spergel} et~al.(2015){Spergel}, {Gehrels} et~al.}]{Spergel15}
{Spergel}, D., {Gehrels}, N., {Baltay}, C., et~al., 2015, arXiv e-prints,
  arXiv:1503.03757, \eprint arXiv:{1503.03757}

\bibitem[{{Stanishev} et~al.(2018){Stanishev}, {Goobar} et~al.}]{Stanishev18}
{Stanishev}, V., {Goobar}, A., {Amanullah}, R., et~al., 2018, \aap, 615, A45

\bibitem[{{Steeghs} et~al.(2022){Steeghs}, {Galloway} et~al.}]{Steeghs22}
{Steeghs}, D., {Galloway}, D.~K., {Ackley}, K., et~al., 2022, \mnras, 511, 2,
  2405, \eprint arXiv:{2110.05539}

\bibitem[{{Steeghs} et~al.(2020){Steeghs}, {Kotak} et~al.}]{Steeghs20oms}
{Steeghs}, D., {Kotak}, R., {Galloway}, D.~K., et~al., 2020, Transient Name
  Server Discovery Report, 2020-2090, 1

\bibitem[{{Stritzinger} et~al.(2011){Stritzinger}, {Phillips}
  et~al.}]{Stritzinger11}
{Stritzinger}, M.~D., {Phillips}, M.~M., {Boldt}, L.~N., et~al., 2011, \aj,
  142, 5, 156, \eprint arXiv:{1108.3108}

\bibitem[{{Sullivan} et~al.(2010){Sullivan}, {Conley} et~al.}]{Sullivan10}
{Sullivan}, M., {Conley}, A., {Howell}, D.~A., et~al., 2010, \mnras, 406, 2,
  782, \eprint arXiv:{1003.5119}

\bibitem[{{Thorp} \& {Mandel}(2022)}]{Thorp22}
{Thorp}, S., {Mandel}, K.~S., 2022, \mnras, 517, 2, 2360, \eprint
  arXiv:{2209.10552}

\bibitem[{{Thorp} et~al.(2021){Thorp}, {Mandel}, {Jones}, {Ward} \&
  {Narayan}}]{Thorp21}
{Thorp}, S., {Mandel}, K.~S., {Jones}, D.~O., {Ward}, S.~M., {Narayan}, G.,
  2021, \mnras, 508, 3, 4310, \eprint arXiv:{2102.05678}

\bibitem[{{Tonry} et~al.(2018){Tonry}, {Denneau} et~al.}]{ATLAS}
{Tonry}, J.~L., {Denneau}, L., {Heinze}, A.~N., et~al., 2018, \pasp, 130, 988,
  064505, \eprint arXiv:{1802.00879}

\bibitem[{{Tripp}(1998)}]{Tripp98}
{Tripp}, R., 1998, \aap, 331, 815

\bibitem[{{Tucker} et~al.(2022){Tucker}, {Shappee} et~al.}]{Tucker22}
{Tucker}, M.~A., {Shappee}, B.~J., {Huber}, M.~E., et~al., 2022, \pasp, 134,
  1042, 124502, \eprint arXiv:{2210.09322}

\bibitem[{{Uddin} et~al.(2020){Uddin}, {Burns} et~al.}]{Uddin20}
{Uddin}, S.~A., {Burns}, C.~R., {Phillips}, M.~M., et~al., 2020, \apj, 901, 2,
  143, \eprint arXiv:{2006.15164}

\bibitem[{{Van Der Walt} et~al.(2011){Van Der Walt}, {Colbert} \&
  {Varoquaux}}]{numpy11}
{Van Der Walt}, S., {Colbert}, S.~C., {Varoquaux}, G., 2011, Computing in
  Science \& Engineering, 13, 22, \eprint arXiv:{1102.1523}

\bibitem[{{Ward} et~al.(2022){Ward}, {Thorp} et~al.}]{Ward22}
{Ward}, S.~M., {Thorp}, S., {Mandel}, K.~S., et~al., 2022, arXiv e-prints,
  arXiv:2209.10558, \eprint arXiv:{2209.10558}

\bibitem[{{Weyant} et~al.(2018){Weyant}, {Wood-Vasey} et~al.}]{Weyant18}
{Weyant}, A., {Wood-Vasey}, W.~M., {Joyce}, R., et~al., 2018, \aj, 155, 5, 201,
  \eprint arXiv:{1703.02402}

\bibitem[{{Williams} et~al.(2020){Williams}, {Hook} et~al.}]{Williams20}
{Williams}, S.~C., {Hook}, I.~M., {Hayden}, B., et~al., 2020, \mnras, 495, 4,
  3859, \eprint arXiv:{2005.07112}

\bibitem[{{Wood-Vasey} et~al.(2008){Wood-Vasey}, {Friedman}
  et~al.}]{Wood-Vasey08}
{Wood-Vasey}, W.~M., {Friedman}, A.~S., {Bloom}, J.~S., et~al., 2008, \apj,
  689, 1, 377, \eprint arXiv:{0711.2068}

\bibitem[{{Wyatt}(2021)}]{Wyatt21pfs}
{Wyatt}, S., 2021, Transient Name Server Classification Report, 2021-2003, 1

\bibitem[{{Yang} et~al.(2019){Yang}, {Sand} et~al.}]{Yang19}
{Yang}, S., {Sand}, D.~J., {Valenti}, S., et~al., 2019, \apj, 875, 1, 59,
  \eprint arXiv:{1901.08474}

\end{thebibliography}

\setlength\LTleft{0pt}
\setlength\LTright{0pt}
\begin{longtable*}[!hbt]{@{\extracolsep{\fill}} lcrr}
\caption{Optical-only SALT3 fit results}\label{tab:SALT3_fit_params}\\

\toprule
SN & Optical Peak MJD & SALT3 $x_1$ & SALT3 $c$ \\
 & (days) & & \\
\hline

2020fxa & 58958.6 $\pm$ 0.3 & 0.89 $\pm$ 0.22 & $-$0.07 $\pm$ 0.03 \\
2020jdo & 58984.1 $\pm$ 0.3 & $-$1.84 $\pm$ 0.37 & 0.18 $\pm$ 0.08 \\
2020jfc & 58985.2 $\pm$ 0.1 & $-$2.90 $\pm$ 0.19 & 0.34 $\pm$ 0.05 \\
2020jgl & 58993.0 $\pm$ 0.3 & $-$0.67 $\pm$ 0.40 & $-$0.00 $\pm$ 0.04 \\
2020jht & 58990.8 $\pm$ 0.2 & 1.17 $\pm$ 0.27 & 0.46 $\pm$ 0.03 \\
2020jio & 58993.6 $\pm$ 0.3 & 5.00 $\pm$ 0.63 & 0.45 $\pm$ 0.05 \\
2020jjf & 58988.2 $\pm$ 0.3 & 0.16 $\pm$ 0.30 & 0.04 $\pm$ 0.04 \\
2020jjh & 58993.4 $\pm$ 0.2 & 0.35 $\pm$ 0.18 & 0.02 $\pm$ 0.03 \\
2020jsa & 58992.9 $\pm$ 0.2 & $-$1.89 $\pm$ 0.19 & 0.28 $\pm$ 0.04 \\
2020jwl & 58992.4 $\pm$ 0.3 & $-$0.25 $\pm$ 0.27 & 0.00 $\pm$ 0.04 \\
2020kav & 58989.1 $\pm$ 0.6 & $-$1.69 $\pm$ 0.62 & $-$0.08 $\pm$ 0.06 \\
2020kaz & 58992.5 $\pm$ 0.2 & 0.56 $\pm$ 0.28 & $-$0.19 $\pm$ 0.04 \\
2020kbw & 58997.1 $\pm$ 0.3 & 0.02 $\pm$ 0.36 & 0.02 $\pm$ 0.06 \\
2020kcr & 58998.2 $\pm$ 0.3 & 5.00 $\pm$ 0.04 & 0.35 $\pm$ 0.06 \\
2020khm & 58993.0 $\pm$ 0.4 & $-$2.20 $\pm$ 0.61 & $-$0.06 $\pm$ 0.08 \\
2020kkc & 58998.5 $\pm$ 0.2 & $-$0.06 $\pm$ 0.28 & 0.12 $\pm$ 0.05 \\
2020kku & 58995.3 $\pm$ 0.3 & $-$2.43 $\pm$ 0.37 & 0.34 $\pm$ 0.07 \\
2020kpx & 59004.7 $\pm$ 0.1 & $-$1.68 $\pm$ 0.11 & $-$0.10 $\pm$ 0.04 \\
2020kqv & 58997.7 $\pm$ 0.4 & $-$0.25 $\pm$ 0.45 & $-$0.14 $\pm$ 0.05 \\
2020kru & 58997.2 $\pm$ 0.2 & $-$1.19 $\pm$ 0.23 & 0.53 $\pm$ 0.06 \\
2020krw & 58999.1 $\pm$ 0.2 & $-$1.25 $\pm$ 0.36 & 0.15 $\pm$ 0.08 \\
2020kyx & 59008.8 $\pm$ 0.1 & $-$0.53 $\pm$ 0.13 & 0.08 $\pm$ 0.04 \\
2020kzn & 59000.9 $\pm$ 0.3 & 1.20 $\pm$ 0.34 & 0.61 $\pm$ 0.07 \\
2020lfe & 59009.5 $\pm$ 0.2 & 0.40 $\pm$ 0.21 & $-$0.00 $\pm$ 0.04 \\
2020lil & 59011.7 $\pm$ 0.2 & $-$1.68 $\pm$ 0.13 & 0.03 $\pm$ 0.05 \\
2020lsc & 59018.2 $\pm$ 0.2 & $-$0.26 $\pm$ 0.12 & 0.03 $\pm$ 0.03 \\
2020lwj & 59015.4 $\pm$ 0.2 & $-$1.39 $\pm$ 0.21 & $-$0.12 $\pm$ 0.05 \\
2020may & 59025.1 $\pm$ 0.2 & 1.66 $\pm$ 0.50 & $-$0.13 $\pm$ 0.06 \\
2020mbf & 59020.2 $\pm$ 0.2 & $-$0.63 $\pm$ 0.18 & 0.08 $\pm$ 0.04 \\
2020mby & 59022.9 $\pm$ 0.2 & $-$1.43 $\pm$ 0.12 & $-$0.10 $\pm$ 0.05 \\
2020mdd & 59016.9 $\pm$ 0.0 & $-$1.82 $\pm$ 0.00 & $-$0.22 $\pm$ 0.03 \\
2020mnv & 59028.3 $\pm$ 0.1 & $-$0.53 $\pm$ 0.13 & 0.00 $\pm$ 0.03 \\
2020mvp & 59028.0 $\pm$ 0.2 & $-$0.45 $\pm$ 0.22 & 0.51 $\pm$ 0.05 \\
2020naj & 59036.3 $\pm$ 0.2 & 4.32 $\pm$ 0.33 & 0.08 $\pm$ 0.03 \\
2020nbo & 59028.8 $\pm$ 0.2 & $-$2.33 $\pm$ 0.23 & 0.07 $\pm$ 0.06 \\
2020ndv & 59034.3 $\pm$ 0.2 & 0.52 $\pm$ 0.16 & $-$0.08 $\pm$ 0.03 \\
2020ned & 59034.6 $\pm$ 0.2 & 5.00 $\pm$ 0.02 & 0.68 $\pm$ 0.04 \\
2020nef & 59029.7 $\pm$ 0.2 & $-$1.38 $\pm$ 0.28 & $-$0.16 $\pm$ 0.06 \\
2020npb & 59041.7 $\pm$ 0.2 & 0.76 $\pm$ 0.16 & 0.02 $\pm$ 0.03 \\
2020nst & 59031.6 $\pm$ 0.3 & 1.21 $\pm$ 0.36 & 0.47 $\pm$ 0.06 \\
2020nta & 59036.1 $\pm$ 0.3 & $-$4.32 $\pm$ 0.35 & 0.56 $\pm$ 0.13 \\
2020ocv & 59048.4 $\pm$ 0.3 & 0.49 $\pm$ 0.34 & 0.15 $\pm$ 0.05 \\
2020oil & 59054.1 $\pm$ 0.2 & $-$0.06 $\pm$ 0.17 & $-$0.18 $\pm$ 0.04 \\
2020oms & 59050.3 $\pm$ 0.3 & 0.73 $\pm$ 0.25 & 0.00 $\pm$ 0.04 \\
2020pst & 59064.6 $\pm$ 0.2 & $-$0.58 $\pm$ 0.20 & $-$0.10 $\pm$ 0.04 \\
\multicolumn{4}{r}{(\emph{continued on next page})} \\
2020qic & 59071.1 $\pm$ 0.2 & 0.31 $\pm$ 0.18 & $-$0.13 $\pm$ 0.03 \\
2020qne & 59075.1 $\pm$ 0.2 & 5.00 $\pm$ 0.03 & 0.50 $\pm$ 0.05 \\
2020rgz & 59089.4 $\pm$ 0.1 & 0.18 $\pm$ 0.13 & 0.25 $\pm$ 0.03 \\
2020rlj & 59088.7 $\pm$ 0.3 & $-$2.35 $\pm$ 0.23 & 0.18 $\pm$ 0.06 \\
2020sjo & 59107.6 $\pm$ 0.2 & $-$0.57 $\pm$ 0.13 & $-$0.03 $\pm$ 0.02 \\
2020sme & 59112.9 $\pm$ 0.2 & 1.85 $\pm$ 0.21 & 0.08 $\pm$ 0.03 \\
2020svo & 59113.5 $\pm$ 0.1 & 0.45 $\pm$ 0.18 & $-$0.06 $\pm$ 0.03 \\
2020swy & 59116.1 $\pm$ 0.2 & 1.08 $\pm$ 0.18 & 0.16 $\pm$ 0.03 \\
2020szr & 59118.0 $\pm$ 0.1 & 0.39 $\pm$ 0.14 & 0.07 $\pm$ 0.03 \\
2020tdy & 59118.8 $\pm$ 0.3 & $-$0.84 $\pm$ 0.36 & $-$0.05 $\pm$ 0.07 \\
2020tfc & 59114.4 $\pm$ 0.2 & $-$2.41 $\pm$ 0.20 & $-$0.02 $\pm$ 0.04 \\
2020tkp & 59123.9 $\pm$ 0.1 & 1.02 $\pm$ 0.18 & 0.27 $\pm$ 0.03 \\
2020tpf & 59123.6 $\pm$ 0.2 & 1.29 $\pm$ 0.23 & $-$0.13 $\pm$ 0.03 \\
2020tug & 59127.2 $\pm$ 0.2 & 0.19 $\pm$ 0.27 & $-$0.17 $\pm$ 0.04 \\
2020uea & 59126.7 $\pm$ 0.1 & $-$3.97 $\pm$ 0.20 & 0.82 $\pm$ 0.09 \\
2020uec & 59131.5 $\pm$ 0.2 & 0.61 $\pm$ 0.16 & $-$0.08 $\pm$ 0.03 \\
2020uek & 59125.1 $\pm$ 0.1 & $-$1.54 $\pm$ 0.16 & 0.08 $\pm$ 0.04 \\
2020uen & 59125.7 $\pm$ 0.0 & $-$0.62 $\pm$ 0.16 & 0.28 $\pm$ 0.04 \\
2020unl & 59135.0 $\pm$ 0.2 & $-$2.00 $\pm$ 0.17 & 0.18 $\pm$ 0.03 \\
2020vnr & 59142.9 $\pm$ 0.4 & 1.28 $\pm$ 0.55 & $-$0.06 $\pm$ 0.06 \\
2020vwv & 59148.4 $\pm$ 0.2 & 0.73 $\pm$ 0.22 & $-$0.05 $\pm$ 0.04 \\
2020wcj & 59148.8 $\pm$ 0.2 & $-$1.30 $\pm$ 0.15 & $-$0.09 $\pm$ 0.03 \\
2020wgr & 59148.3 $\pm$ 0.2 & 0.17 $\pm$ 0.19 & 0.08 $\pm$ 0.04 \\
2020wtq & 59148.8 $\pm$ 0.1 & $-$0.76 $\pm$ 0.17 & $-$0.04 $\pm$ 0.04 \\
2020yjf & 59164.6 $\pm$ 0.3 & 0.59 $\pm$ 0.23 & $-$0.29 $\pm$ 0.05 \\
2020ysl & 59162.5 $\pm$ 0.3 & $-$1.39 $\pm$ 0.20 & $-$0.00 $\pm$ 0.06 \\
2020aczg & 59217.4 $\pm$ 0.2 & $-$0.04 $\pm$ 0.17 & $-$0.13 $\pm$ 0.03 \\
2021ash & 59245.7 $\pm$ 0.3 & $-$0.35 $\pm$ 0.21 & $-$0.19 $\pm$ 0.04 \\
2021aut & 59245.0 $\pm$ 0.3 & 0.03 $\pm$ 0.27 & $-$0.11 $\pm$ 0.04 \\
2021bbz & 59251.9 $\pm$ 0.4 & $-$0.17 $\pm$ 0.17 & $-$0.01 $\pm$ 0.03 \\
2021biz & 59258.4 $\pm$ 0.1 & $-$0.42 $\pm$ 0.13 & 0.02 $\pm$ 0.02 \\
2021bjy & 59250.5 $\pm$ 0.3 & $-$2.05 $\pm$ 0.24 & $-$0.10 $\pm$ 0.05 \\
2021bkw & 59251.8 $\pm$ 0.0 & $-$1.60 $\pm$ 0.14 & 0.04 $\pm$ 0.03 \\
2021ble & 59258.3 $\pm$ 0.3 & 1.37 $\pm$ 0.32 & $-$0.08 $\pm$ 0.04 \\
2021dnm & 59281.3 $\pm$ 0.2 & 1.05 $\pm$ 0.26 & 0.13 $\pm$ 0.05 \\
2021fof & 59299.5 $\pm$ 0.2 & $-$0.28 $\pm$ 0.18 & $-$0.00 $\pm$ 0.04 \\
2021fxy & 59306.1 $\pm$ 0.2 & 0.28 $\pm$ 0.14 & $-$0.07 $\pm$ 0.03 \\
2021ghc & 59307.3 $\pm$ 0.3 & 0.46 $\pm$ 0.23 & $-$0.06 $\pm$ 0.04 \\
2021glz & 59307.5 $\pm$ 0.2 & 3.00 $\pm$ 0.28 & 0.15 $\pm$ 0.03 \\
2021hiz & 59320.7 $\pm$ 0.1 & $-$0.09 $\pm$ 0.10 & $-$0.14 $\pm$ 0.02 \\
2021huu & 59323.1 $\pm$ 0.2 & 0.37 $\pm$ 0.19 & 0.00 $\pm$ 0.04 \\
2021J & 59233.4 $\pm$ 0.2 & 0.22 $\pm$ 0.10 & $-$0.25 $\pm$ 0.03 \\
2021lug & 59353.5 $\pm$ 0.2 & 0.85 $\pm$ 0.26 & $-$0.27 $\pm$ 0.03 \\
2021mim & 59362.1 $\pm$ 0.2 & $-$1.95 $\pm$ 0.17 & 0.11 $\pm$ 0.05 \\
2021pfs & 59391.9 $\pm$ 0.1 & $-$0.55 $\pm$ 0.11 & 0.02 $\pm$ 0.03 \\
2021usd & 59442.9 $\pm$ 0.1 & 0.59 $\pm$ 0.12 & 0.17 $\pm$ 0.03 \\
2021zfq & 59488.2 $\pm$ 0.3 & $-$3.29 $\pm$ 0.24 & $-$0.40 $\pm$ 0.20 \\
2021zfs & 59484.4 $\pm$ 0.2 & $-$1.46 $\pm$ 0.15 & $-$0.13 $\pm$ 0.07 \\
2021zfw & 59483.7 $\pm$ 0.1 & $-$2.51 $\pm$ 0.20 & $-$0.47 $\pm$ 0.13 \\

\bottomrule

\end{longtable*}

\end{document}